\documentclass[acmtog, authorversion]{acmart}

\AtBeginDocument{%
  \providecommand\BibTeX{{%
    \normalfont B\kern-0.5em{\scshape i\kern-0.25em b}\kern-0.8em\TeX}}}

\acmDOI{xxxx}

\acmJournal{FACMP}
\acmVolume{0}
\acmNumber{0}
\acmArticle{0}
\acmMonth{0}

\usepackage{latexsym}
\usepackage{url}
\usepackage[utf8]{inputenc}
\usepackage{caption}
\usepackage{subcaption}
\usepackage[ruled,vlined]{algorithm2e}
\usepackage{xcolor}
\usepackage{times}
\usepackage{epsfig}
\usepackage{amsmath}
\usepackage{multirow}
\usepackage{enumitem}
\usepackage{float}
\usepackage{graphicx}

\begin{document}

\title{GraphSeam: Supervised Graph Learning Framework for Semantic UV Mapping}

\author{Fatemeh Teimury}
\affiliation{\institution{McGill University}~and~\institution{Autodesk}\country{Canada}}
\email{fatemeh.teimury@mail.mcgill.ca}

\author{Bruno Roy}
\affiliation{\institution{Autodesk}\country{Canada}}
\email{bruno.roy@autodesk.com}

\author{Juan Sebastián Casallas}
\affiliation{\institution{Autodesk}\country{USA}}
\email{sebastian.casallas@autodesk.com}

\author{David MacDonald}
\affiliation{\institution{Autodesk}\country{Canada}}
\email{david.macdonald@autodesk.com}

\author{Mark Coates}
\affiliation{\institution{McGill University}\country{Canada}}
\email{mark.coates@mcgill.ca}

\renewcommand{\shortauthors}{Teimury et al.}

\begin{abstract}
Recently there has been a significant effort to automate UV mapping, the process of mapping 3D-dimensional surfaces to the UV space while minimizing distortion and seam length. Although state-of-the-art methods, Autocuts and OptCuts, addressed this task via energy-minimization approaches, they fail to produce semantic seam styles, an essential factor for professional artists. The recent emergence of Graph Neural Networks (GNNs), and the fact that a mesh can be represented as a particular form of a graph, has opened a new bridge to novel graph learning-based solutions in the computer graphics domain.
In this work, we use the power of supervised GNNs for the first time to propose a fully automated UV mapping framework that enables users to replicate their desired seam styles while reducing distortion and seam length. To this end, we provide augmentation and decimation tools to enable artists to create their dataset and train the network to produce their desired seam style. 
We provide a complementary post-processing approach for reducing the distortion based on graph algorithms to refine low-confidence seam predictions and reduce seam length (or the number of shells in our supervised case) using a skeletonization method.
\end{abstract}

\begin{CCSXML}
<ccs2012>
   <concept>
       <concept_id>10010147.10010371.10010396</concept_id>
       <concept_desc>Computing methodologies~Shape modeling</concept_desc>
       <concept_significance>500</concept_significance>
       </concept>
   <concept>
       <concept_id>10010147.10010257.10010293</concept_id>
       <concept_desc>Computing methodologies~Machine learning approaches</concept_desc>
       <concept_significance>500</concept_significance>
       </concept>
 </ccs2012>
\end{CCSXML}

\ccsdesc[500]{Computing methodologies~Shape modeling}
\ccsdesc[500]{Computing methodologies~Machine learning approaches}

\keywords{graph neural networks, UV mapping, semantic, post-processing}

\begin{teaserfigure}
    \centering
    \includegraphics[height=225px]{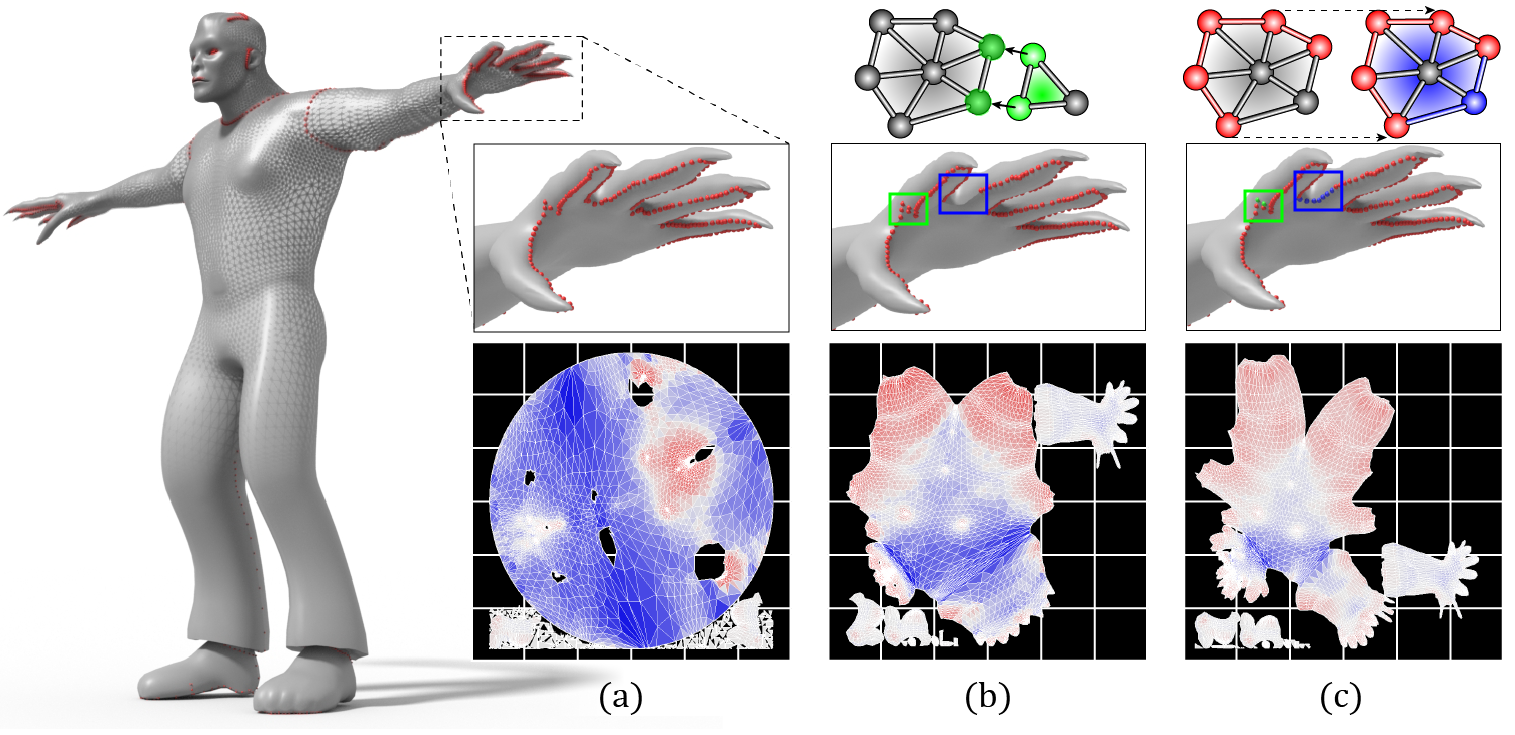}
    \label{fig:approach-teaser}
    \caption{The proposed approach on a sample mesh. In the top row, we show the predicted outputs of our graph network in UV space (a). Additional post-processing steps (bottom row) improves our results by removing small shells (b) and by adding missing seams (c) reducing shell count and distortion.}
\end{teaserfigure}

\maketitle

\section{Introduction}

UV mapping is a fundamental task in computer graphics that involves the projection of 3D surfaces to 2D representations. This process requires unwrapping the 3D mesh from the surface for texture and colour assignment.
Although different evaluation metrics exist for UV Mapping, prior methods ~\cite{poranne2017autocuts,li2018optcuts,sheffer2002seamster} solely focus on optimizing two critical metrics: distortion and seam length. The main difference between earlier and more recent methods for UV Mapping is that the earlier methods, such as~\cite{sheffer2002seamster}, use separate steps for optimizing seam length and distortion since these parameters have different natures (distortion is continuous, and seam length is discrete). In contrast, recent methods \cite{poranne2017autocuts,li2018optcuts} employ iterative frameworks to optimize seam length and distortion jointly.

Although these methods have demonstrated promising performance that minimizes distortion and seam length, they usually produce a single shell. Producing UV maps with semantic boundaries is a critical goal, and to our knowledge, this goal has not been addressed so far.
Moreover, artists usually ask for a framework that can mimic similar seam styles on objects within the same category. For example, the seam style for a collection of humanoid models should be consistent. Ideally, the framework should also  spend the same amount of time for each model.

Polygonal meshes have many intrinsic similarities with graphs, and this motivated us to build our proposed method using state-of-the-art graph learning approaches. Graph-based neural network architectures (e.g., graph convolution networks and graph attention networks) have been widely used recently in the machine learning community \cite{kipf2016semi,velivckovic2017graph} and have proved promising for analysis of data on graphs and point clouds. Our proposed method represents the first application of such learning techniques for seam detection in the UV Mapping context. Moreover, the supervised nature of these algorithms enables the proposed method to reproduce similar styles on test objects, a feature that artists are always looking for. In addition, the translation of the distortion minimization task to the graph learning context allows us to propose a post-processing algorithm based on the Steiner tree problem \cite{sheffer2002seamster} and skeletonization~\cite{abu2013skeletonization,youssef2015graph,yang2019novel}. The post-processing step helps to produce seams that align with semantic boundaries and reduces the number of shells. 

Although we demonstrate our framework's efficiency using a specific dataset (Autodesk\textsubscript{\textregistered}Character Generator), we appreciate that there is a great variation between 3D models and  seam style based on the dataset type. On the other hand, the artist's task is usually to define a particular seam style over a large set of objects, optimistically taking months to complete. So, we provide augmentation and decimation tools to enable artists to train the proposed network with any type of objects or seam style they require. We suggest that artists apply their desired seam style on a few models manually and then use our augmentation tools to create their dataset. The main motivation for producing seam cuts manually is that defining optimal cuts and seam style depends on multiple factors such as curvature, distortion, and the UV space. Providing manually labeled examples means that artists will generate a superior training dataset for graph networks. In summary, our main contributions are:

\begin{enumerate}[leftmargin=*]
    \itemsep0em
    \item{we leverage the supervised nature of graph learning methods to mimic semantic seam style in UV Mapping;}
    \item{we incorporate dual graphs and state-of-the-art graph-based learning methods to address seam detection as an edge-classification task;}
    \item{we suggest key informative edge features for our framework which provide a considerable improvement in all 3D mesh edge-based networks;}
    \item{we restate the distortion minimization problem in UV Mapping in terms of graph connectivity; and}
    \item{we then propose a post-processing procedure based on the Steiner Tree problem~\cite{sheffer2002seamster} and skeletonization to improve seam detection network results.}
\end{enumerate}

\section{Related Work}\label{Related}
Surface parametrization is the projection of a surface to 2D space and is central to a broad spectrum of problems in the computer graphics and animation communities. 
Surface parametrization can be addressed via two separate pipelines: (i) specifying seams and (ii) minimizing distortion~\cite{julius2005d, khodakovsky2003globally,sheffer2002seamster,sander2007multi,desbrun2002intrinsic}. Initial UV mapping methods either considered only one of the problems or considered both but addressed them via two sequential pipelines. Recent methods have started to focus on addressing minimizing distortion and seam detection simultaneously \cite{poranne2017autocuts,li2018optcuts}. In  this  section, we  briefly  review  initial  methods  that  address one of the tasks of specifying seams \cite{julius2005d,khodakovsky2003globally,sheffer2002seamster,sander2007multi} or minimizing distortion \cite{desbrun2002intrinsic,khodakovsky2003globally} and then focus on the state-of-the-art methods that simultaneously address both \cite{poranne2017autocuts,li2018optcuts}.

Specifying cuts on the surface is the initial step to reach the surface parametrization. The papers \cite{julius2005d,khodakovsky2003globally,sheffer2002seamster,sander2007multi}  propose different architectures that specify seams as the initial step and then minimize distortion by adding more cuts. None of these frameworks is capable of preserving semantic boundaries automatically; they rely on guidance from a user. The aim of parametrization approaches that strive to minimize distortion is to preserve angles and areas in order to produce isometric mappings\cite{khodakovsky2003globally, desbrun2002intrinsic,liu2008revisiting,sander2001texture}. The main limitation of all of these pipelines is that the nonlinearity of the optimization function is not satisfactorily addressed.

The difference between the nature of distortion and cuts (i.e., distortion is continuous and cuts are discrete) has hampered the design of an architecture that optimizes both parameters at the same time. So, in all the above-mentioned methods, the parametrization is mainly dependent on specifying cuts and then if the cuts do not produce the desired distortion value, the architecture needs to suggest a new set of cuts. Recently, \cite{poranne2017autocuts} proposed Autocuts, which is a fully automated architecture that minimizes distortion and optimizes cuts using an energy-based solver that is capable of iteratively and jointly converging on soft constraints such as distortion metrics. Later, \cite{li2018optcuts} proposed the OptCuts algorithm that starts from an arbitrary initial embedding and satisfies the distortion bounds requested by the user.

The main deficiency of these methods is that the final cuts do not preserve semantic boundaries. Although \cite{poranne2017autocuts, li2018optcuts} simultaneously optimize distortion and cuts, they generate results with very few shells (e.g., a single shell for our Character Generator dataset) while losing semantic boundaries. 
Another drawback of the state-of-the-art methods is the mesh resolution scalability. Based on our experiments, meshes with more than 2000k polygons require extensive resources. State-of-the-art methods construct a matrix of high-resolution meshes that requires a huge amount of memory and leads to expensive computations throughout the operation of their iterative solvers. Eventually, it is desirable to incorporate these algorithms in interactive tools used by artists and designers; due to the very high computational and memory requirements of existing techniques, the proposed methods struggle to be adequately responsive. In addition to these two limitations, neither Autocuts nor OptCuts guarantee convergence and whether meaningful results are obtained or not depends heavily upon the initializations. In other words, without very carefully chosen manual inputs from the user, even if the algorithms do reach convergence, the end-results might be unusable in practice.

\section{Problem Statement}\label{Problem Statement}

Our main goal is to address UV mapping for 3D meshes. 
Since different evaluation metrics exist for UV mapping such as distortion (ratio of 2d area to 3d area or the ratio of 2d perimeter to 3d perimeter for each face), the number of UV shells, semantic boundaries, 2d layout efficiency (what percentage of the UV space is occupied), we address our problem via two different pipelines. In the first pipeline, we focus on finding a good initial seam placement, and in the second, we refine the seams to minimize the distortion.

\subsection{Seam detection}\label{Seam detection}
We have access to a set of 3D meshes, $M_{i} = (V_{i},E_{i}, F_{i})$ $i \in 1,2,\dots,K$ with $\left | V_{i} \right|= N_{i} $ nodes. $E_{i} \in V_{i}\times V_{i}$ denotes the set of edges, and $F_{i} \in V_{i}\times V_{i}\times V_{i}$ denotes the set of triangle faces for the $i$-th mesh. Let $X_{i} \in R^{N\times f}$ be the node feature matrix where $f$ is the number of features of every node. $A_{i} \in R^{N_{i}\times N_{i}}$ is the adjacency matrix corresponding to the original mesh, and $L_{i} \in (0,1)^{1\times E_{i}}$ is the edge label vector for all edges in the mesh $i$. A value of $0$ indicates that the edge is not a seam and $1$ indicates that it is.
In our setting, we have a labelled training set and a test set where both the meshes and labels of the test set are unavailable during training. The task is to predict all edge labels for meshes in the test set.

\subsection{Distortion Minimization}\label{Distortion}
Each mesh is defined as $M = (V, E, F) $ where $V$ is a set of nodes (in our context these are vertices), $E$ is a set of edges, and $F$ is a set of faces in the mesh. Also, we have access to a vector $L \in (0,1)^{1\times \left | E \right |}$ which corresponds to proposed edge labels (i.e., seam or non-seam), and a vector of face distortions $D_{f} \in R ^{ 1\times \left | F \right |}$ in which each element is a distortion value for a face of the mesh $M$ in the UV space. The task is to produce new edge labels ${L}' \in R^{1\times \left | E \right |}$ that represent a modification, including both addition of missing seams and removal of spurious seams, to the initial edge labels $L$, with the goal of reducing distortion in the shells.

\section{Background}\label{Background}

\subsection{Graph Neural Networks (GNNs)}\label{sec:gnn}

In the past years, there has been intensive research into the development of neural networks that can be applied to graph-structured data\cite{levie2018cayleynets,defferrard2016convolutional,henaff2015deep,estrach2014spectral}. At a high level, the core difference between conventional neural networks and GNNs is the need for more flexibility in the convolution (or aggregation) operations that take place at each layer. Since graphs are irregular and there is no ordering or position associated with each node, the aggregation operator should be invariant to the ordering of the nodes in the neighbourhood. Also, since nodes have different degrees, the size of the neighbourhoods vary.
We now review the GNNs that we have experimented with in our proposed framework.

\paragraph{Graph Convolutional Network (GCN)}\label{background gcn}
The GCN~\cite{kipf2016semi} is one of the simpler GNN approaches. Let $\hat{A}_G = D^{-1/2}(A+I)D^{-1/2}$ be a normalized adjacency matrix, where $I$ is the identity matrix, $A$ is the original adjacency matrix of the graph $G$, and $D$ is the degree matrix. Let $\sigma$ be a non-linear activation function and denote by $W^{(k)}$ the neural network weights at layer $k$. Denoting the output of layer $k$ as $H^{(k+1)}$ and letting $X$ be the input feature matrix, the graph convolution operation for a GCN can be written as follows:
\begin{equation}
\begin{split}
    H^{(1)} = \sigma \left ( \widehat{A}_{G} XW^{(0)} \right ) \\
    H^{(k+1)} = \sigma \left ( \widehat{A}_{G} H^{(k)}W^{(k)} \right ) 
\end{split}
\label{gcn operation}
\end{equation}

\paragraph{Graph Attention Networks (GAT)}\label{background gat} While GCN \cite{kipf2016semi} uses a simple mean aggregation function, a GAT\cite{velivckovic2017graph} adopts an attention mechanism to learn how much weight a node should place on another node in its neighbourhood when performing the aggregation. The graph convolution operation in a GAT can be expressed for node $v$ as:
\begin{equation}
h_{v}^{(k)} = \sigma \left (\sum_{u \in N(v)\cup v} \alpha_{uv}^{(k)}W^{(k)}h_{u}^{(k-1)}\right) \,,
\end{equation}
where $N(v)$ is the (possibly multi-hop) neighborhood of node $v$, $h_{v}^{(0)} = x_v$ is the input feature vector, and  $\alpha_{uv}^{(k)}$ is the attention weight that node $v$ associates with node $u$. The attention weight can be calculated via the equation:
\begin{equation}
\alpha_{uv}^{(k)} = softmax\left ( g\left ( a^{T}\left [ W^{(k)}h_{v}^{(k-1)}\left |  \right |W^{(k)}h_{u}^{(k-1)} \right ] \right ) \right ).
\end{equation}
where $[\cdot||\cdot]$ denotes concatenation of two vectors, $g$ is the Leaky RELU activation function and $ a $ is the vector of learnable parameters. The $softmax$ function is used for attention weight normalization. There are other possible choices for constructing the attention function. Multi-head attention can be used to improve prediction performance.

\paragraph{GraphSAGE}\label{background graphsage} 
Many deeper GNNs suffer from over smoothing (aggregation occurs over too large a neighborhood). The size of a multi-hop neighborhood can expand exponentially, and this can make training for very large graphs extremely slow. GraphSAGE~\cite{hamilton2017inductive} addresses these two issues by using residual (skip) connections to prevent oversmoothing and employing neighbor sampling to place limits on the size of the neighborhood used for computations. The method allows for a variety of  different aggregation mechanisms, including concatenation, max-pooling, and LSTMs. We can summarize the GraphSAGE structure by:
\begin{equation}
    h_{v}^{(k)} = \sigma (W^{k}\cdot f_{(k)}(h_{v}^{(k-1)},\left \{ h_{u}^{(k-1)},\forall u \in S_{N(v)} \right \}))\,.
\end{equation}
Here $h_{v}^{0} = x_v$ is the input feature vector for node $v$, $f_{k}(\cdot)$ is the aggregation function and $S_{N(v)}$ is the set of sampled neighbors of the node $v$.

\paragraph{Graph Isomorphism Network (GIN)}\label{background gin} In ~\cite{xu2018powerful}, Xu et al.\ highlight the inability of common GNNs such as the GCN and GAT to learn different node embeddings to distinguish between different graph structures. Xu et al.\ demonstrate that a maximally powerful graph neural network can be constructed by using a multi-layer perceptron for the the aggregation combined with an irrational scalar. This GNN, called a Graph Isomorphism Network (GIN), is maximally powerful in the sense that if any GNN can distinguish between two different graphs, then GIN is also capable of distinguishing between them. Note that this power is not equivalent to performance on learning tasks such as node or edge classification. Denote by $\epsilon^{(l)}$ the irrational scalar associated with layer $l$ and let $MLP(\cdot)$ represent a multi-layer perceptron. The aggregation step in GIN can be written as: 
\begin{equation}
    h_{v}^{(k)} = MLP(( 1+\epsilon^{(k)})h_{v}^{\left ( k-1 \right )} +\sum_{u \in N\left ( v \right )}h_{u}^{\left ( k-1 \right )})\,.
\end{equation}

\subsection{Dual Graph Convolutional Neural Networks}\label{background dual gcnn}
In this section, we first introduce dual graphs and then we provide a brief explanation of methods that generalize graph neural networks via dual graphs.

\begin{figure*}[t]
    \centering
    \includegraphics[width=\linewidth]{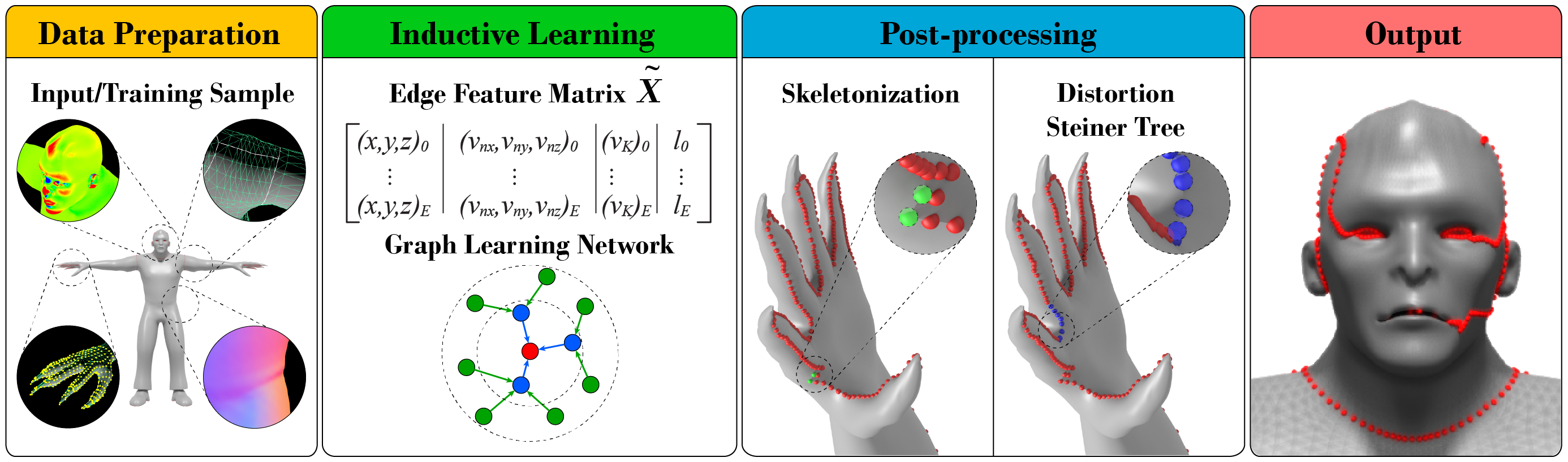}
    \caption{Pipeline overview of the proposed approach.}
    \label{fig:approach-overview}
\end{figure*}

\paragraph{Dual Graphs} Let $G = (V,E)$ be the original undirected graph. The dual graph (also known as line graph) of $G$, denoted by $\widetilde{G} = (\widetilde{V} = E , \widetilde{E})$, is constructed such that every vertex of the dual graph $\widetilde{v}$ corresponds to an edge $(i,j)\in E$ in the original graph $G$. Consider any two dual vertices $\tilde{v}=(i,j)$ and $\tilde{v}' =({i}',{j}') \in \widetilde{V}$ in the dual graph $\widetilde{G}$. If these are connected by an edge in dual graph, then the corresponding edges $(i,j)$ and $(i',j')$ must share an endpoint in the original graph $G$.
The definition can be extended to directed graphs, but we focus on the undirected case because there is no direction associated with 3D mesh edges. 

The GNN architectures can be directly applied to the dual graph or can be extended to learn jointly over both the original and dual graphs~\cite{zhuang2018dual,monti2018dual,zhang2019dual,jepsen2019graph}. Application on the dual graph is beneficial when the goal is classification or regression of edges. Joint learning can improve performance because the architecture can more readily learn structural relationships between both edges and nodes. Since one of our tasks is classifying edges as seams, it is natural to perform learning on the dual graph. 

\subsection{Steiner Tree\cite{sheffer2002seamster}}\label{background steiner tree}

A Steiner tree for a set of vertices (terminals) in a graph is a connected sub-graph containing all the terminal vertices~\cite{skiena1998algorithm}. A minimal Steiner tree is a Steiner tree with minimal sum of edge weights.
The problem of finding the minimal Steiner tree is NP-Complete \cite{skiena1998algorithm}. The following is a standard algorithm to approximate the minimal Steiner tree. It has been proven to be within a factor of $\frac{2}{\sqrt{3}}$ of the optimum\cite{skiena1998algorithm}.

\begin{enumerate}
    \itemsep0em
    \item  For each $n,m \in T$ where $T$ is the tree, compute the shortest path $P(n,m)$ between vertices $n$ and $m$.
    \item Define a new graph where $T$ are the vertices and there is an edge between each pair of vertices. The new weight for each edge is set to the weight of the shortest path between the two terminals.
    \item Compute the minimal spanning tree on the new graph.
\end{enumerate}

\section{Methodology}\label{Methodology}
Our proposed method consists of two separate blocks: seam detection and minimizing distortion and thinning seam lines. A more detailed pipeline of our approach is shown in Figure~\ref{fig:approach-overview}.

\subsection{Graph Learning for Seam Detection}\label{Graph Learning Seam Detection}

Polygonal meshes have many intrinsic similarities with graphs, and indeed, can be fully specified by an annotated graph. With such a representation, the seam detection task can be expressed as edge classification.

State-of-the-art graph learning methods such as GAT and GCN focus primarily on addressing node or graph classification. On the other hand, UV mapping aligns more naturally with edge classification. Instead of using the feature matrix and adjacency of the original graph, in which every element corresponds to a node, we construct an edge feature matrix and consider the dual graph.

Our suggested framework is flexible and can be applied to any edge feature matrix, but we propose a specific edge feature matrix in Section~\ref{Dataset features}. We choose edge features that we have found experimentally to be particularly useful when trying to discriminate between seams and non-seams.

\subsection{Edge Classification}\label{Dataset features}
When performing edge classification, we first create a node feature matrix using normalized vertex coordinates $(x,y,z)$, vertex normals $(v_{nx},v_{ny},v_{nz})$ obtained from the object content, and discrete vertex Gaussian curvature $v_K$, a vital distortion related feature that has been used in~\cite{sheffer2002seamster}. To prepare the edge feature matrix $\tilde{X}$, we used concatenation of the two endpoints node features for each edge. For example, $\tilde{x}_{\tilde{v}=(i,j)} = [x_i||x_j]$ where $x_{i}$ is the node feature corresponding to node $i$.

A concern with the strategy outlined above is that during the concatenation procedure, we must choose the ordering of the two nodes that form an edge. Since there is no ordering in the graph and  edges are undirected, this can lead to ambiguity. To resolve this, we follow the procedure of~\cite{monti2018dual} and construct an augmented dual graph. Each original edge is mapped to two nodes in the augmented dual graph. For an edge $(i,j)$ in the original graph, the feature vectors associated with these two dual nodes are $[x_i||x_j]$ and $[x_j||x_i]$, respectively. Both nodes are connected to all nodes that share a common endpoint in the original graph ($i$ or $j$). During experiments, we observe that this process, although desirable in terms of ensuring a unique representation, has minimal effect on performance. We therefore report the results achieved by assigning a random ordering to the nodes and using the standard dual. Because the constructed graph is smaller and sparser, the computational demands are considerably less, which is important for large meshes.  

After constructing the dual graph, we apply a graph neural network, choosing the aggregation strategy that best matches the characteristics of the seam detection task. In particular, we find that it is critical to avoid over-smoothing and the loss of local information. For this reason, we incorporate residual connections after every layer in all of the GNNs we employ\cite{he2016deep,li2019deepgcns,li2019fi,huang2019residual,zhou2020effective}. 

\subsection{Distortion Minimizing Steiner Tree}\label{Distortion Steiner Tree}

After applying the seam detection procedure, we have a vector of edge probabilities $L_{p} \in R^{1 \times \left | E \right | }$ where $\left | E \right |$ is the number of edges. These probabilities are the softmax outputs of the GNN, and indicate how likely an edge is to belong to a seam. We can apply a threshold to these probabilities to derive binary classifications. The threshold determines the number of connected components that will be derived, and should be adjusted according to the size of the mesh and the desired number of shells.

We do not provide explicit encouragement to the GNN  to produce contiguous seams, although there is implicit encouragement through the graph aggregation process. As a result, seams can be incomplete, missing a handful of critical edges. We address this problem by considering the graph connectivity. First, we separate the original graph into connected components by performing cuts along the edges labelled as seams. In each connected component, there is a path between any two nodes. 

After this step, we reason that within one of these connected component, all remaining seam edges should be connected --- a seam is supposed to define a boundary and there should not be isolated, incomplete seams. We are thus motivated to construct a tree that connects all of the nodes (in the original graph) that have been identified as belonging to one or more seam edges. We formulate this task as a Steiner tree problem, assigning edge weights in order to reduce the distortion. We then use the approximate algorithm presented in Section~\ref{background steiner tree} to derive the tree. The edges identified in this tree are considered to be the new seams. 

In order to reduce the distortion, we define the edge weights based on the face distortion in the UV space. We cut the mesh using the edge labels produced by the seam detection network, then construct the UV map and calculate the distortion value for every face in the mesh. The normalized distortion value for a face is the ratio of face area in UV space to the face area in 3D space, assuming that UV space has been scaled so that the total area of all faces matches that of 3D space. As a result, a face distortion value of 1 corresponds to no distortion, values higher than one correspond to areas of magnification in UV space, and values less than 1 correspond to minification. We define the face distortion vector as $D_{F} \in R^{1 \times \left| F \right|}$ where $ \left| F \right|$ is the number of faces in the mesh $m$. Every edge $e$ in a manifold triangulated mesh is incident on two faces. We define the edge weight as follows:
\begin{equation}\label{eq:edge_length}
l_{e} =  1-\left| D_{f1,e} -  D_{f2,e}\right|
\end{equation}
where $D_{f1,e}$ and $D_{f2,e}$ correspond to the normalized face distortion values of the first and second face neighbors of the edge $e$ (the ordering is irrelevant since we are taking the absolute value of the difference). If there is a significant difference between the distortion values for the two faces, then the edge is a a good candidate for a seam. The assigned weight is low, so the approximate Steiner tree algorithm is encouraged to include them when constructing the seam tree.

\paragraph{Skeletonization} 

The seam prediction network often produces thick regions of several candidate edges connected to each other, whereas we require seams to be topologically 1-dimensional curves. To address this, we incorporate an additional step that refines and thins the seam detection predictions. We use a relatively straightforward adaptation of the idea of skeletonization of graphs and images~\cite{abu2013skeletonization,youssef2015graph,yang2019novel}. This first requires the estimated edge probabilities to be converted to a probability value per vertex, which is simply the maximum probability of any edge incident on the vertex. The user supplies a threshold to define the set of candidate vertices which will be thinned. A value between 10\% to 30\% is typically a good selection, and the particular choice will affect the topology of the seams somewhat, with higher values producing more loops in the result. The skeletonization then proceeds by repetitively removing the lowest probability vertex of this set whose removal does not change the connectivity of the set of vertices, which corresponds to the essence of the image-based skeletonization algorithms. In order to avoid reducing long thin regions to a single vertex at the centre, a vertex is not allowed to be removed if it would then make the distance from one of the deleted vertices to the remaining vertices higher than a user-defined threshold. This threshold is typically set to the distance of about two to four edges. In this way, the essential loops and branches of the candidate seams are preserved, but guaranteed to be exactly one edge thick. The skeletonization step also explicitly eliminates tiny shells that only include one or two triangles, using another user-defined threshold.

\begin{figure}[h]
    \centering
    \includegraphics[width=\linewidth]{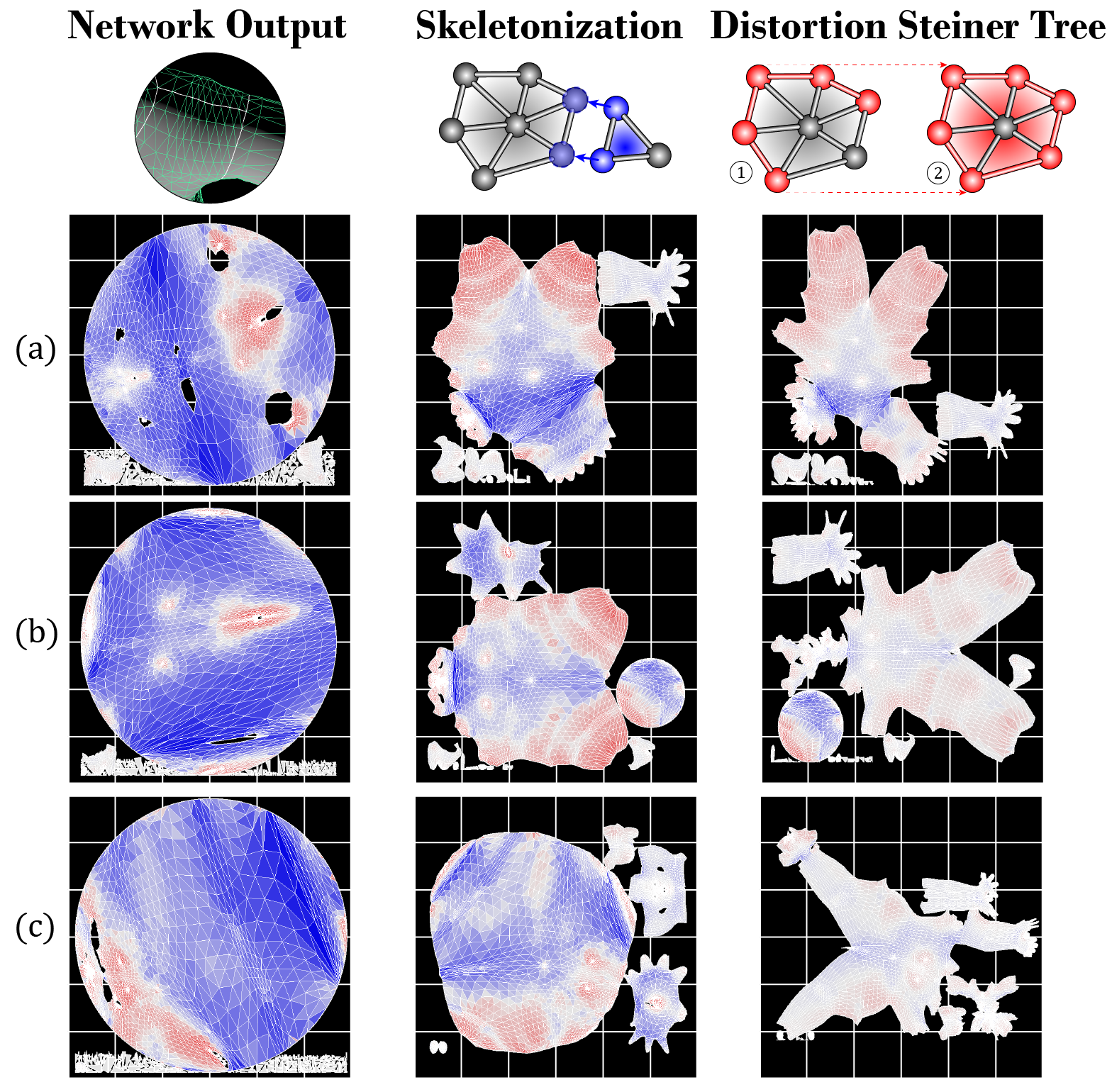}
    \caption{We present the effect of the proposed post-processing steps on the UV shells of three (3) test 3D models, respectively from top to bottom: Ken, Sibilla, and Xena. The UV shells are color-coded to describe the resulting distortion at each step where blue shows compression and red stretching.}
    \label{fig:approach-post-processing}
\end{figure}

\section{Experimental Results}\label{Experimental Results}
\subsection{Dataset}\label{Dataset}

We conducted experiments on the Autodesk\textsubscript{\textregistered}~Character Generator (CG) dataset~\cite{CharacterGenerator}. CG dataset consists of procedurally generated 3D humanoid models, and includes seam labels generated using a specific seam style. Technical artists within Autodesk\textsubscript{\textregistered} have provided ground truth seams for the 3D models. Also available are geometric features such as vertex normals and vertex coordinates. The dataset is provided in commonly used file formats for data (e.g., OBJ and PLY). The CG dataset consists of $93$ training objects and $3$ validation objects and $3$ test objects. 
We decimated all the objects to 10000 faces resolution using the Autodesk\textsubscript{\textregistered}~decimation tool, and in the remainder of the paper, we will refer to this decimated dataset as CG10000.
We provide a visualization of the resulting UV shells of the test objects in Figure~\ref{fig:approach-post-processing}. In addition, we employ an augmentation tool to produce many augmented meshes. This tool creates new meshes by adding Gaussian random noise to the coordinates of random vertexes  of the original mesh. In this setting, users can choose the number of augmented objects and the mean and variance of the Gaussian random noise.
Since we are conducting supervised inductive learning for UV mapping for the first time, there is no benchmark dataset currently available. Autodesk\textsubscript{\textregistered}~Character Generator is available for users, allowing for experimentation with the dataset.

\subsection{Baseline Methods}

\begin{figure}[t]
    \centering
    \includegraphics[width=\linewidth]{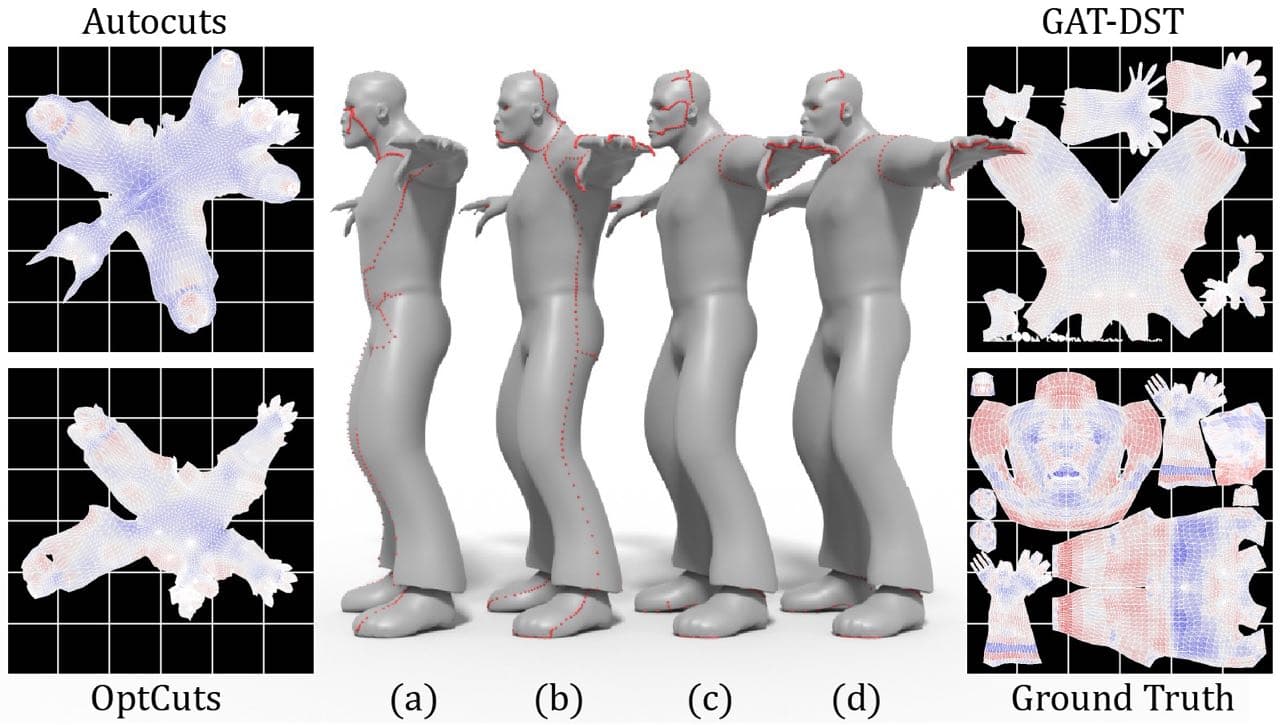}
    \caption{Comparing our GAT-based method with DST refinement (c) with Autocuts (a), OptCuts (b), and ground truth (d).}
    \label{fig:approach-compare-STAR}
\end{figure}

We compare the performance of our approach with state-of-the-art energy-based UV Mapping methods, Autocuts~\cite{poranne2017autocuts} and OptCuts~\cite{li2018optcuts}. We compare to these algorithms in two ways. First, we apply them directly on the 3D models (i.e., without any seam initialization). This allows us to evaluate how well our proposed method compares to the state-of-the-art as an automated tool for UV mapping. We compare quantitatively using the distortion metric and the number of generated shells. We also conduct a qualitative assessment of whether the methods are able to identify semantic boundaries. Secondly, we evaluate how well Autocuts and OptCuts perform as post-processing methods to improve the quality of seams. We apply our proposed seam detection method to provide an initial seam specification. This allows us to compare to our proposed Distortion Steiner Tree algorithm.

\subsection{Proposed Methods}

We experimented with a selection of graph neural networks, employing modified versions of GCN~\cite{kipf2016semi}, GAT~\cite{velivckovic2017graph}, GraphSAGE (GS)~\cite{hamilton2017inductive}, and GIN~\cite{xu2018powerful}. We used the inductive learning implementation of GAT provided by the Deep Graph Library (DGL)~\cite{wang2019deep} and we implemented inductive versions of the rest of the baselines using DGL similar to an inductive GAT implementation to provide a fair comparison. Moreover, we add skip connections to every layer of each architecture. The task of seam detection is binary classification, but the classes are imbalanced (fewer than 10\% of the edges are seam edges). We therefore employ a weighted cross-entropy loss, using weight $100$ for seams and $1$ for non-seam edges.

\paragraph{Hyperparameters} We used grid search for hyper-parameter tuning, using the validation set of 3 objects. The search grids are specified in the supplementary material. For each GNN architecture we use 3 layers with 64 hidden units per layer. In GAT~\cite{velivckovic2017graph} structure, the number of hidden attention units is 3, the number of output attention units is 5, and the attention drop is 0.2. For GraphSAGE~\cite{hamilton2017inductive}, we use an LSTM aggregator. Results for other GraphSAGE aggregators are reported in the supplementary material. GIN~\cite{xu2018powerful} has two extra MLP layers with 64 hidden units. For all the models, the early stop threshold is 50 and the learning rate is $0.0005$. Learning is conducted using the Adam optimizer.

\subsection{Results and Discussion}

\paragraph{GNN-based Seam Detection} Table~\ref{table:CG_10000_pef_metrics} compares the performance of the proposed method with different GNNs to Autocuts and OptCuts. These methods do not perform seam detection  in the supervised manner as our proposed method, but instead perform an optimized mapping after seams have been specified by a user. In this table, we compare to the single-shell solution derived by these algorithms. 

\begin{table}[t]
\centering
\caption{Performance of seam detection and UV mapping. Seam detection is evaluated using false positive rate (FPR), true positive rate (TPR) and accuracy (Acc.). UV maps are evaluated based on average distortion and number of shells both before post-processing (BPP) and after (APP).}
\vspace{0.1cm}
\label{table:CG_10000_pef_metrics}
\footnotesize
\begin{tabular}{|p{1.6cm}|p{0.4cm}|p{0.5cm}|p{0.5cm}|p{0.5cm}|p{0.5cm}|p{0.4cm}|p{0.4cm}|}
 \hline
  \multirow{2}{*}{\textbf{Method}} & \multicolumn{3}{c|}{\textbf{Perf. Metrics (\%)}} & \multicolumn{2}{c|}{\textbf{Avg. Dist.}} & \multicolumn{2}{c|}{\textbf{\# Shells}} \\ \cline{2-8}
  & \textbf{FPR} & \textbf{TPR} & \textbf{Acc.} & \textbf{BPP} & \textbf{APP} & \textbf{BPP} & \textbf{APP}\\
 \hline
 \hline
 Ground Truth &-&-&-&0.294&-&10&-\\
 \hline
 OptCuts &-&-&-&0.107&-&1&-\\
 \hline
  Autocuts &-&-&-& 0.282&-&1&-\\
 \hline
    Prop-GCN &3.62& 87.36& 96.01 & 1.493&0.310&280&385\\
 \hline    
  Prop-GAT & 0.34 & \textbf{99.00}& 99.63 &0.524 & 0.135& 30.6 &68\\
  \hline
    Prop-GS &\textbf{0.04}&98.56& \textbf{99.90}&0.424&0.125&9&36\\
  \hline
  Prop-GIN &0.85&71.74&98.05&3.829& 0.371 &36&89 \\
 \hline
\end{tabular}
\end{table}

\begin{figure*}[t]
    \centering
    \includegraphics[width=\linewidth]{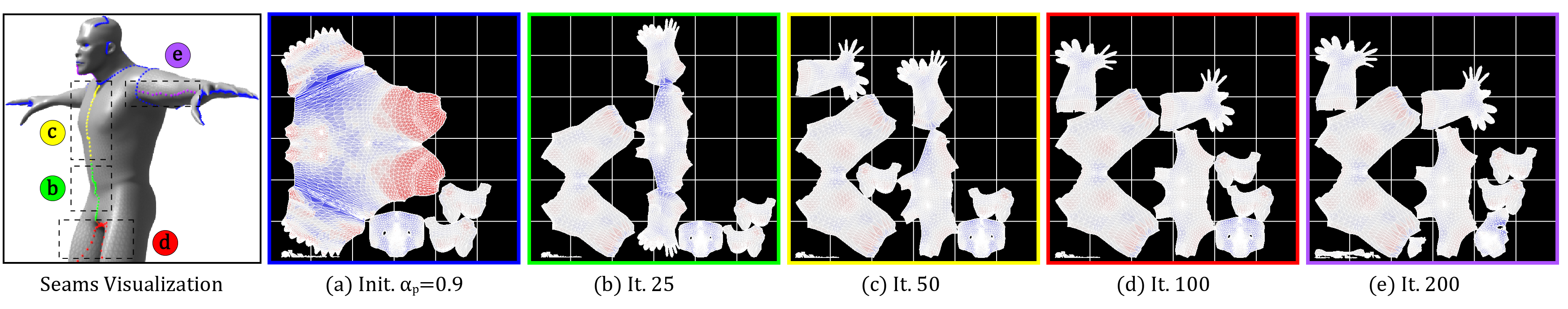}
    \caption{Evolution of UV maps using Autocuts for post-processing, with initialization by GAT-based seam detection method.}
    \label{fig:gat-autocuts}
\end{figure*}

The seam detection results suggest that the simple averaging aggregation of the GCN is inadequate for the task. The attention mechanism of GAT, the LSTM aggregator of GraphSAGE, and the MLP of GIN provide the learning power to derive much better performance. The methods employing GraphSAGE and GAT perform best. In terms of UV mapping, OptCuts produces the lowest distortion map. OptCuts provides a lower distortion map than any of the proposed GNN-based algorithms before application of the post-processing Distortion Steiner Tree algorithm. After applying this, the GAT and GraphSAGE maps provide similar distortion values to the OptCuts map. 

More importantly, unless seams are provided by a user, the energy-based baselines produce solutions with a single shell. The specification of seams is one of the most burdensome components of UV mapping for an artist, so both Autocuts and OptCuts are missing critical functionality. By contrast, the proposed GNN-based methods identify multiple shells and do a much better job of identifying semantic boundaries (consistent with those recognized by the artists providing ground-truth). Figure~\ref{fig:approach-compare-STAR} provides an example of the quality of the maps produced by the approaches.

Prior to the post-processing step, the proposed algorithms fail to separate some of the larger shells identified in the ground truth. The Distortion Steiner Tree algorithm accomplishes the separation, both reducing the distortion and producing smaller shells that more closely match those identified by the artists in the ground truth. The disadvantage is that the method leads to many more single shells.

\begin{table}[b]
\centering
\caption{Comparison of our post-processing method with baselines Autocuts and OptCuts. We report average distortion and number of shells metrics before (BPP) and after post-processing (APP). Autocuts and OptCuts are initialized with the best GNN seam detection result based on GAT. Our proposed method is initialized with GNN seam detection using GAT, GCN, and GraphSAGE (GS) and applies the Distortion Steiner Tree algorithm (DST). The last row shows the effect of Skeletonization (SK).}
\vspace{0.1cm}
\label{table:post_processing}
\footnotesize
\begin{tabular}{|p{1.4cm}|p{1.1cm}|p{0.5cm}|p{0.5cm}|p{0.5cm}|p{0.8cm}|}
 \hline
  \multirow{2}{*}{\textbf{Method}} & \multirow{2}{*}{\textbf{Step(s)}} & \multicolumn{2}{c|}{\textbf{Avg. dist.}} & \multicolumn{2}{c|}{\textbf{\# Shells}}\\ \cline{3-6}
  & & \textbf{BPP} & \textbf{APP} & \textbf{BPP} & \textbf{APP} \\
 \hline
 \hline
 \multirow{4}{*}{Autocuts} & it.~25 & 0.524 & 0.443 & 30.6 & 33\\ \cline{2-6}
  & it.~50 & 0.524 & 0.422 & 30.6 & 35.3\\ \cline{2-6}
  & it.~100 & 0.524 & 0.382 & 30.6 & 35\\ \cline{2-6}
  & it.~200 & 0.524 & 0.404 & 30.6 & 36.3\\
   \hline
 \multirow{4}{*}{OptCuts} & it.~25 & 0.524 & 0.453 & 30.6 & 31.3\\ \cline{2-6}
 & it.~50  & 0.524 & 0.435 & 30.6 & 34.3\\ \cline{2-6}
 & it.~100 & 0.524 & 0.377 & 30.6 & 35\\ \cline{2-6}
 & it.~200 & 0.524 & 0.322 & 30.6 & 37\\
 \hline
 Prop-GAT & DST & 0.524 & 0.135 & 30.6 & 68\\
 \hline
 Prop-GS & DST & 0.424 & \textbf{0.125}& 9 & 35.6\\
 \hline
 Prop-GCN & DST & 1.493 & 0.310 & 280.3 & 385\\
 \hline
 Prop-GCN & DST~\vline~\textbf{SK} & 1.493 & 0.271 & 280.3 & \textbf{19.3}\\
\hline
\end{tabular}
\end{table}

\paragraph{Distortion Steiner Tree Algorithm} Table~\ref{table:post_processing} examines post-processing performance, comparing our proposed Distortion Steiner Tree algorithm to Autocuts and OptCuts (when these are used to refine the initial seams produced by our GAT-based seam detection procedure). The proposed algorithm leads to UV maps with considerably lower distortion.

Fig.~\ref{fig:gat-autocuts} depicts the evolution of the UV maps when using Autocuts for post-processing. Autocuts performs iterative unwrapping operations (unfolding and optimizing the UV layout). We provide initial boundaries using the most probable seams from the graph-learning seam detection (i.e., using a threshold of $\alpha_p\ge0.9$). By iteration 25, Autocuts has separated the upper body from the thighs, and after iterations 50 and 100, the energy-based solver has managed to separate the arms from the upper body. After iteration 100 there are few modifications, although eventually the head is separated from the neck.

Fig.~\ref{fig:approach-compare-STAR}(c) shows the UV map derived by using the GAT-based seam detection followed by the Distortion Steiner Tree algorithm. Comparing this to the map derived by post-processing with Autocuts in Fig.~\ref{fig:gat-autocuts}, we see that the proposed GAT-DST method manages to produce a single shell for the torso, following the ground-truth provided by the artists. In contrast, Autocuts divides the torso into several shells. We observe similar behaviour for multiple examples (please see the supplementary section), indicating that our proposed DST (using face distortion vector information) can better reproduce and preserve artist-specified semantics.

The final row of Table~\ref{table:post_processing} shows the importance of the skeletonization for the GCN-based method, which is more prone to producing thick seams than the other GNNs. The average number of shells is reduced from 280.3 to 19.3. The effect is depicted in the second column of Fig.~\ref{fig:approach-post-processing}. 

 \section{Conclusion and Discussion}\label{Conclusion and Discussion}
We have proposed a novel methodology for the task of UV mapping in computer graphics by leveraging graph learning approaches. The proposed technique uses the dual graph and state-of-the-art graph learning frameworks to address seam detection as an edge classification task. In contrast to existing baselines\cite{poranne2017autocuts,li2018optcuts}, the proposed algorithm produces a solution with multiple shells. Visualization of the results suggests that the algorithm manages to mimic the seam style of the training data.

In order to further reduce the distortion and to catch any seams that the GNN-based method failed to detect, we proposed a graph algorithm based on the Steiner Tree \cite{sheffer2002seamster} to minimize the distortion. Our results demonstrate that application of this algorithm during post-processing considerably reduces the distortion of the UV maps. Furthermore, it performs better than application during post-processing of the energy-minimization methods, Autocuts and OptCuts. 

We developed a skeletonization procedure that can reduce the number of small shells that are identified by our GNN-based seam detection procedure. We believe that a promising future direction involves striving towards incorporating the three methods --- seam detection, post-processing distortion reduction, and skeletonization --- in a single learning framework. This would likely involve the introduction of a differentiable distortion proxy in the objective and regularizers that strongly encourage thin, contiguous seams.

\begin{acks}
This work was supported and funded by Autodesk, Mitacs, and McGill University. We would like to thank Autodesk for providing resources and insightful comments during our research. We would also like to thank Hervé Lange, Group Architect for Entertainment Creation Products at Autodesk, who provided insightful directions to the project, and our internal artists, Sabrina Parent and Pierre Picard, for evaluating our results based on their artistic expertise.
\end{acks}

\bibliographystyle{ACM-Reference-Format}
\bibliography{arxiv_paper}

\appendix

\section{Ground truth}
We include some additional results and figures that provide further illustration of the behaviour our proposed approach GraphSeam.

\begin{figure}[h]
     \centering
     \begin{subfigure}[b]{0.15\textwidth}
         \centering
         \includegraphics[width=\textwidth , height=\textwidth]{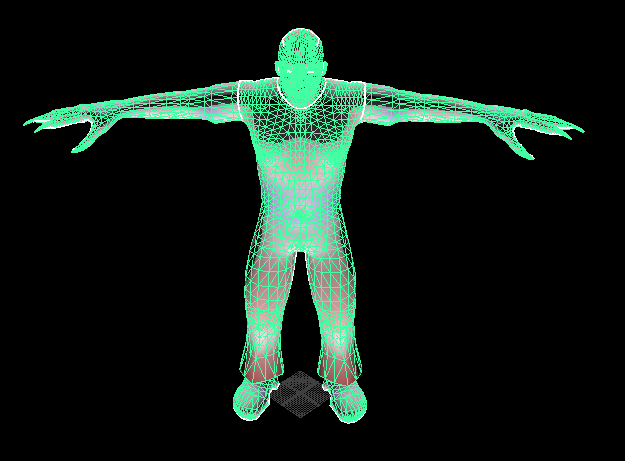}
         \caption{Ken:obj}
         \label{fig:ken_gr}
     \end{subfigure}
     \hfill
      \begin{subfigure}[b]{0.15\textwidth}
         \centering
         \includegraphics[width=\textwidth , height=\textwidth]{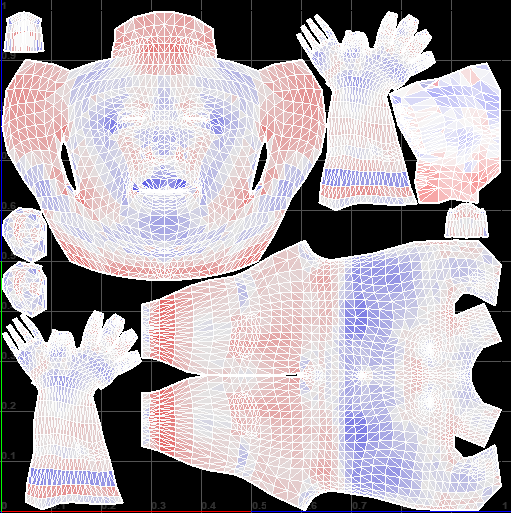}
         \caption{Ken:UV}
         \label{fig:ken_gr}
     \end{subfigure}
     \hfill
     \begin{subfigure}[b]{0.15\textwidth}
         \centering
         \includegraphics[width=\textwidth , height=\textwidth]{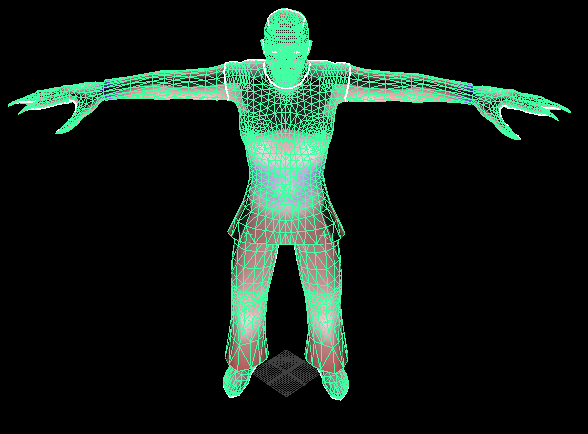}
         \caption{Sibilia:obj}
         \label{fig:sibilia_gr}
     \end{subfigure}
     \hfill
      \begin{subfigure}[b]{0.15\textwidth}
         \centering
         \includegraphics[width=\textwidth , height=\textwidth]{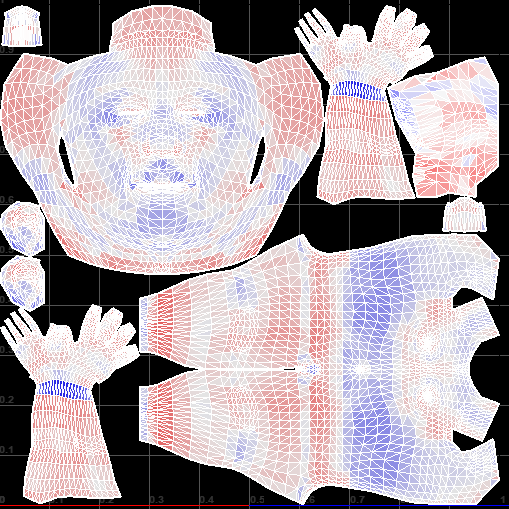}
         \caption{Sibilia:UV}
         \label{fig:sibilia_gr}
     \end{subfigure}
     \hfill
     \begin{subfigure}[b]{0.15\textwidth}
         \centering
         \includegraphics[width=\textwidth , height=\textwidth]{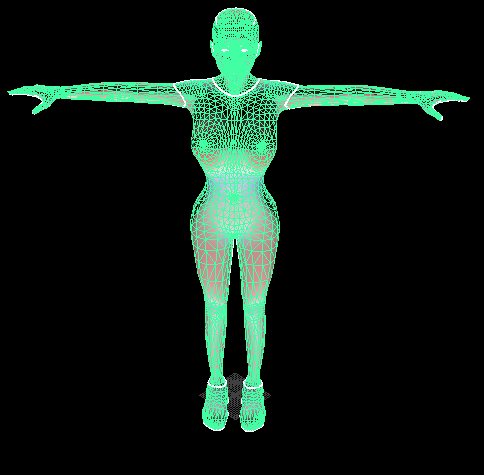}
         \caption{Xena:obj}
         \label{fig:xena_gr}
     \end{subfigure}
     \hfill
     \begin{subfigure}[b]{0.15\textwidth}
         \centering
         \includegraphics[width=\textwidth , height=\textwidth]{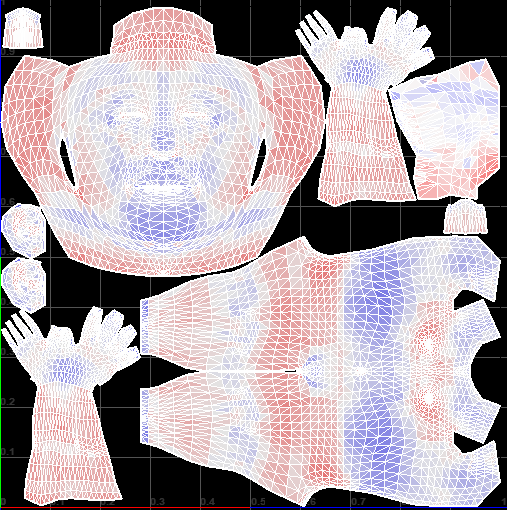}
         \caption{Xena:UV}
         \label{fig:xena_gr}
     \end{subfigure}
        \caption{Visualization of objects and their corresponding UV maps in the test set}
        \label{fig:ground truth}
\end{figure}

In this section we are providing ground truth visualizations (Fig.~\ref{fig:ground truth}) for the test set of the Character Generator 10000 dataset (CG10000).

\section{Augmentation Tool}
\begin{figure}[t]
     \centering
     \begin{subfigure}[b]{0.15\textwidth}
         \centering
         \includegraphics[width=\textwidth , height=\textwidth]{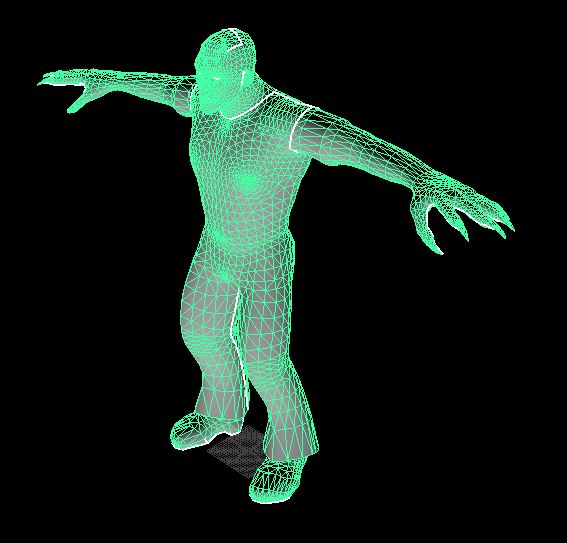}
          \caption{Original model}
         \label{fig:ken_gr}
     \end{subfigure}
     \hfill
      \begin{subfigure}[b]{0.15\textwidth}
         \centering
         \includegraphics[width=\textwidth , height=\textwidth]{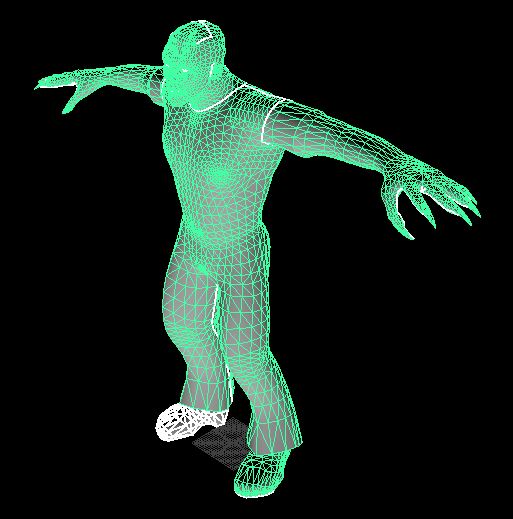}
          \caption{Augmentation 1}
         \label{fig:ken_gr}
     \end{subfigure}
     \hfill
     \begin{subfigure}[b]{0.15\textwidth}
         \centering
         \includegraphics[width=\textwidth , height=\textwidth]{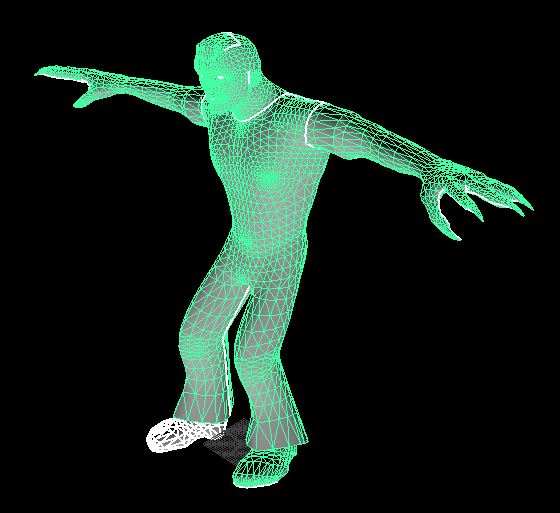}
             \caption{Augmentation 2}
         \label{fig:sibilia_gr}
     \end{subfigure}
     \newline
      \begin{subfigure}[b]{0.15\textwidth}
         \centering
         \includegraphics[width=\textwidth , height=\textwidth]{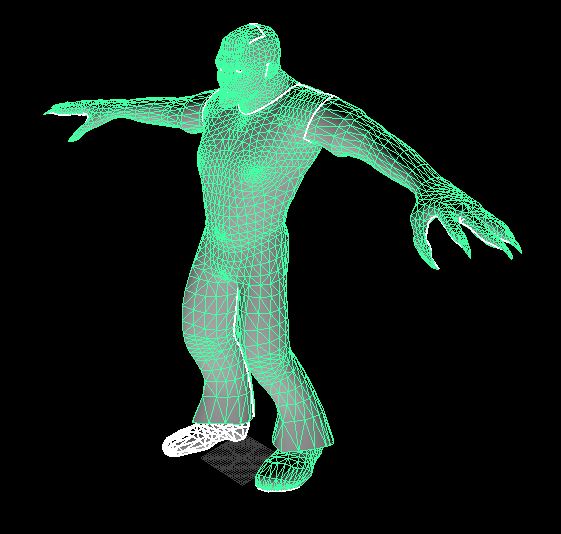}
          \caption{Augmentation 3}
         \label{fig:sibilia_gr}
     \end{subfigure}
     \hfill
     \begin{subfigure}[b]{0.15\textwidth}
         \centering
         \includegraphics[width=\textwidth , height=\textwidth]{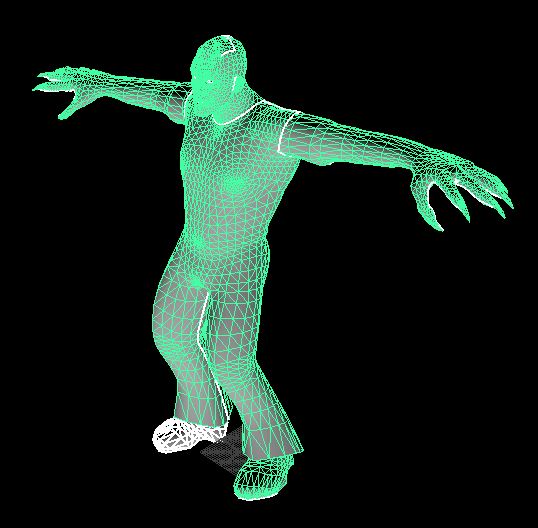}
          \caption{Augmentation 4}
         \label{fig:xena_gr}
     \end{subfigure}
     \hfill
     \begin{subfigure}[b]{0.15\textwidth}
         \centering
         \includegraphics[width=\textwidth , height=\textwidth]{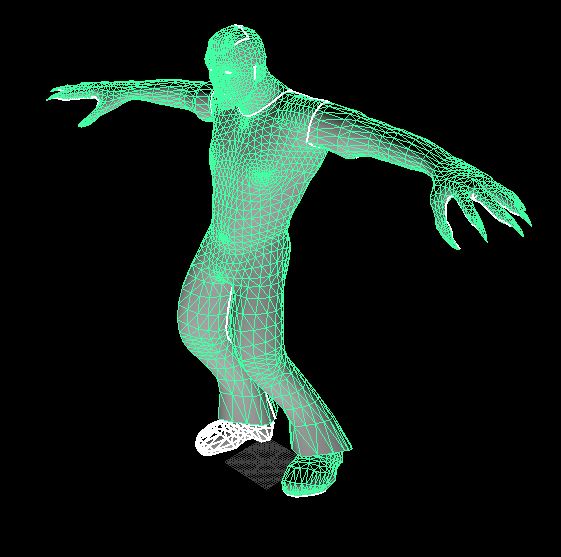}
         \caption{Augmentation 5}
         \label{fig:xena_gr}
     \end{subfigure}
    \caption{Visualization of 3D model Ken and its augmentations.}
    \label{fig:augmentation tool}
\end{figure}

To provide better insight into our proposed augmentation tool, we illustrate five different augmentation visualization for Ken(one of our test models) in Fig.~\ref{fig:augmentation tool}. All the augmentations are produced via adding Gaussian noise that varies vertex positions up to 20\%.

\section{Decimation Tool}
\begin{figure}[h]
     \centering
     \begin{subfigure}[b]{0.15\textwidth}
         \centering
         \includegraphics[width=\textwidth , height=\textwidth]{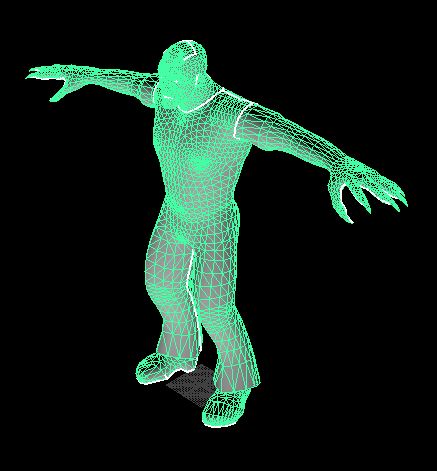}
          \caption{Res:10000f}
         \label{fig:ken_gr}
     \end{subfigure}
     \hfill
      \begin{subfigure}[b]{0.15\textwidth}
         \centering
         \includegraphics[width=\textwidth , height=\textwidth]{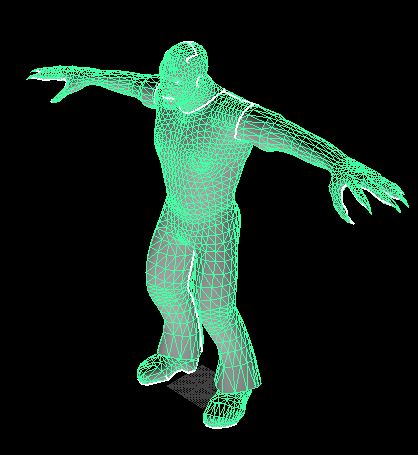}
          \caption{Res:8000f}
         \label{fig:ken_gr}
     \end{subfigure}
     \hfill
     \begin{subfigure}[b]{0.15\textwidth}
         \centering
         \includegraphics[width=\textwidth , height=\textwidth]{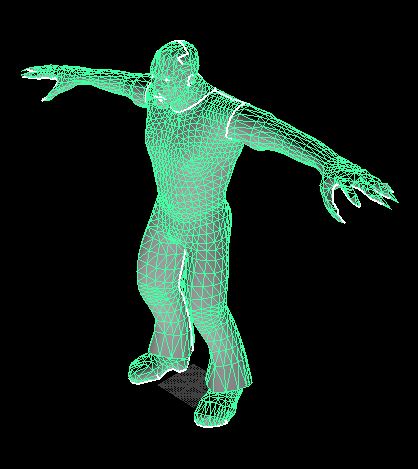}
             \caption{Res:6000f}
         \label{fig:sibilia_gr}
     \end{subfigure}
     \newline
      \begin{subfigure}[b]{0.15\textwidth}
         \centering
         \includegraphics[width=\textwidth , height=\textwidth]{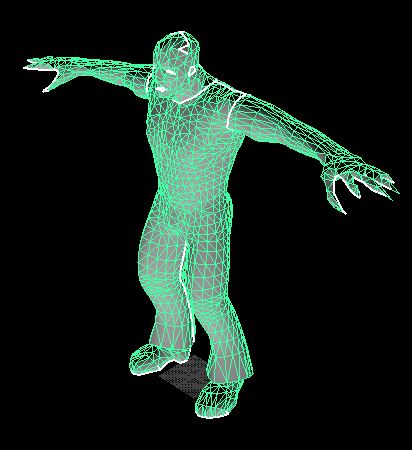}
          \caption{Res:4000f}
         \label{fig:sibilia_gr}
     \end{subfigure}
     \hfill
     \begin{subfigure}[b]{0.15\textwidth}
         \centering
         \includegraphics[width=\textwidth , height=\textwidth]{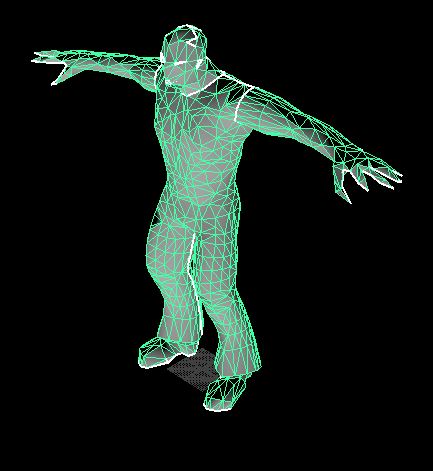}
          \caption{Res:2000f }
         \label{fig:xena_gr}
     \end{subfigure}
     \hfill
     \begin{subfigure}[b]{0.15\textwidth}
         \centering
         \includegraphics[width=\textwidth , height=\textwidth]{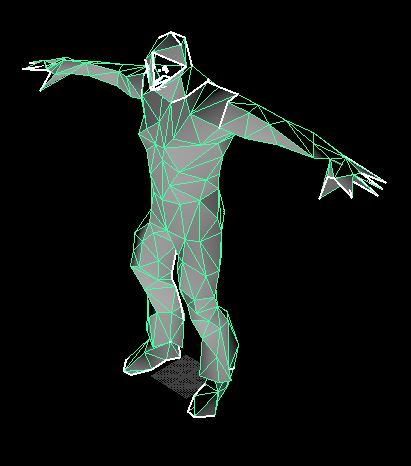}
         \caption{Ken:500 f}
         \label{fig:xena_gr}
     \end{subfigure}
    \caption{Visualization of different resolution of 3D model Ken using decimation tool. Reported resolutions are based on the number of triangulated faces(f).}
    \label{fig:decimation tool}
\end{figure}

To provide better insight into our proposed decimation tool, we illustrate five different resolutions for Ken(one of our test models) in Fig.~\ref{fig:decimation tool}. This tool needs a predefined face resolution and, by removing or adding vertices and edges to the initial object, produces a new object with the predefined resolution. 

\section{Post-processing with Autocuts and Optcuts without initialization}
\begin{figure}[t]
     \centering
     \begin{subfigure}[b]{0.15\textwidth}
         \centering
         \includegraphics[width=\textwidth , height=\textwidth]{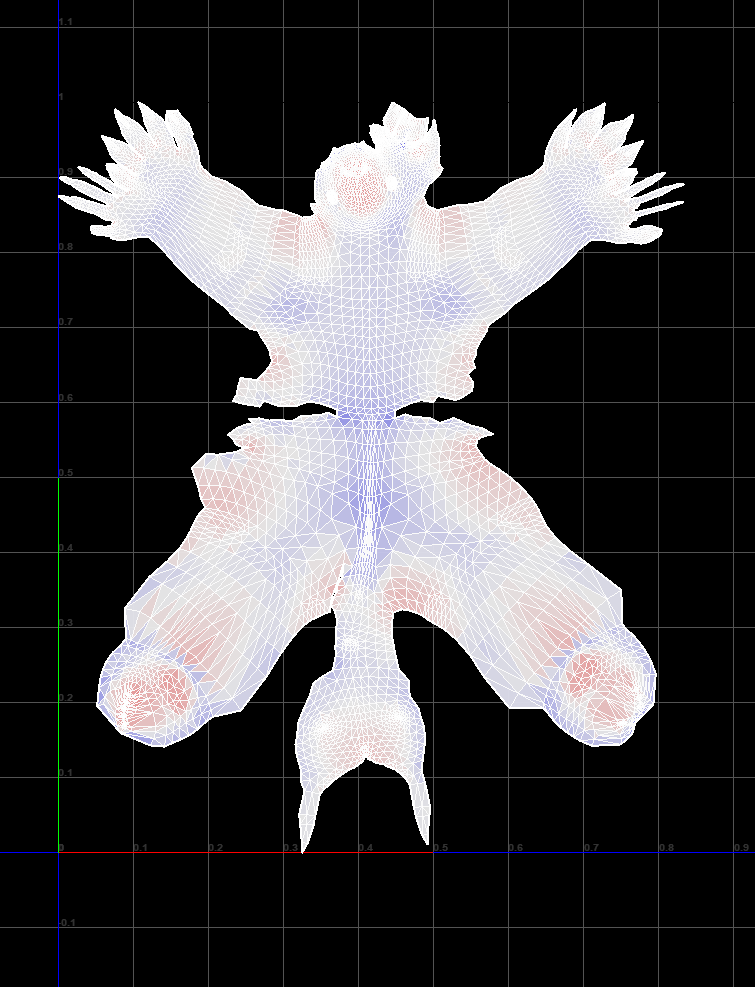}
         \caption{Ken:Autocuts}
         \label{fig:ken_autocuts}
     \end{subfigure}
     \hfill
     \begin{subfigure}[b]{0.15\textwidth}
         \centering
         \includegraphics[width=\textwidth , height=\textwidth]{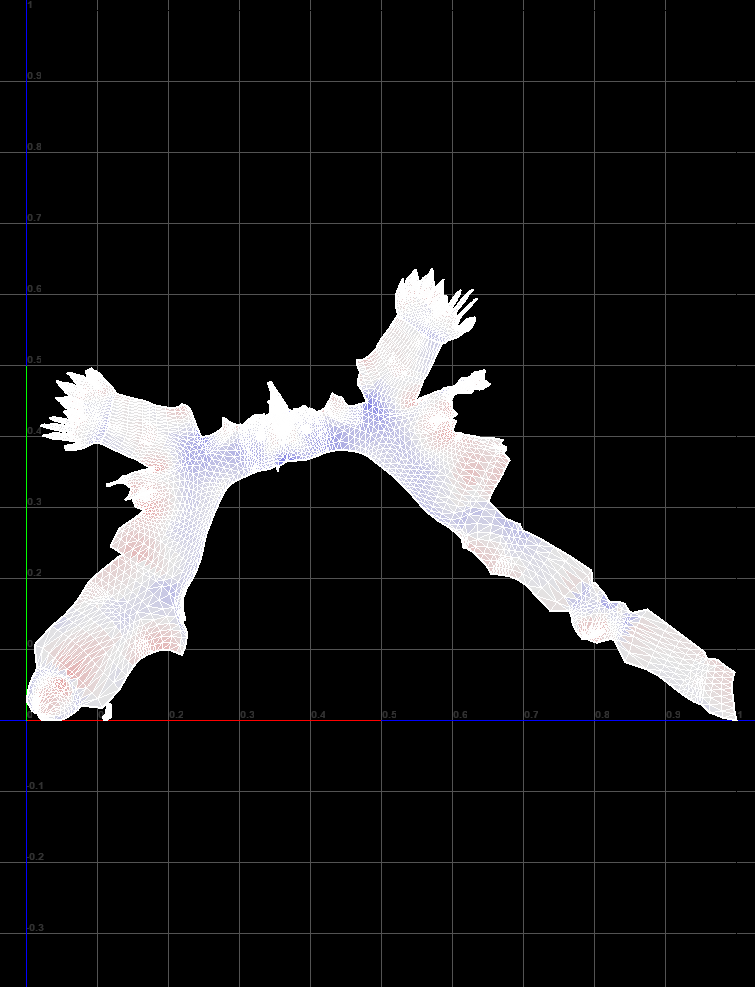}
         \caption{Sibilia:Autocuts}
         \label{fig:sibilia_autocuts}
     \end{subfigure}
     \hfill
     \begin{subfigure}[b]{0.15\textwidth}
         \centering
         \includegraphics[width=\textwidth , height=\textwidth]{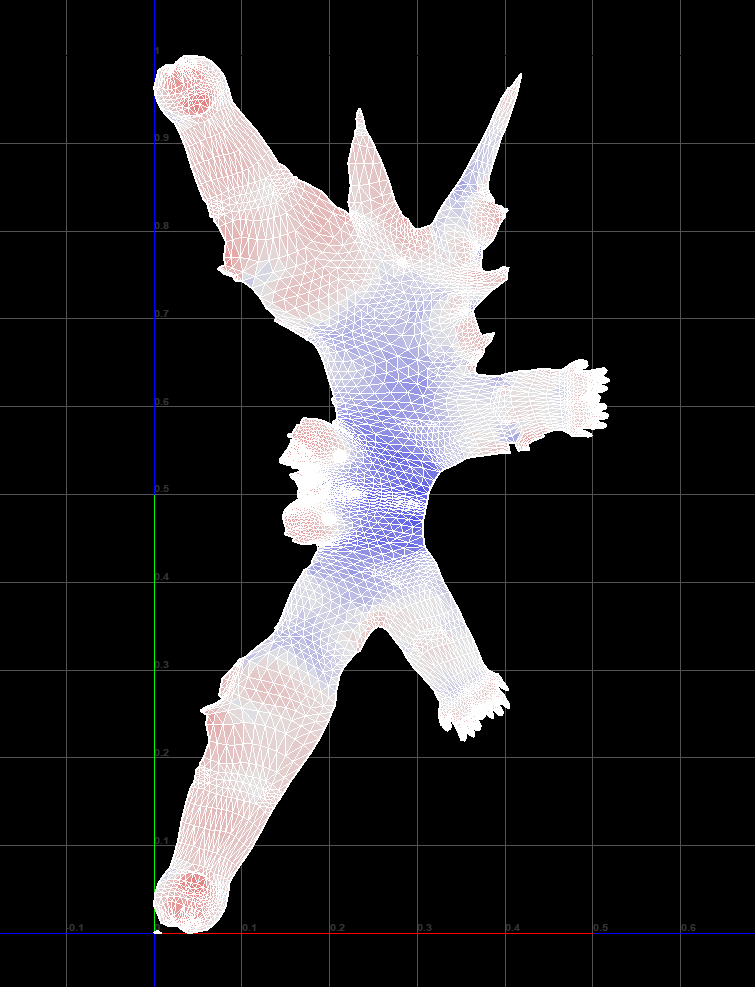}
         \caption{Xena:Autocuts}
         \label{fig:xena_autocuts}
     \end{subfigure}
      \hfill
       \begin{subfigure}[b]{0.15\textwidth}
         \centering
         \includegraphics[width=\textwidth , height=\textwidth]{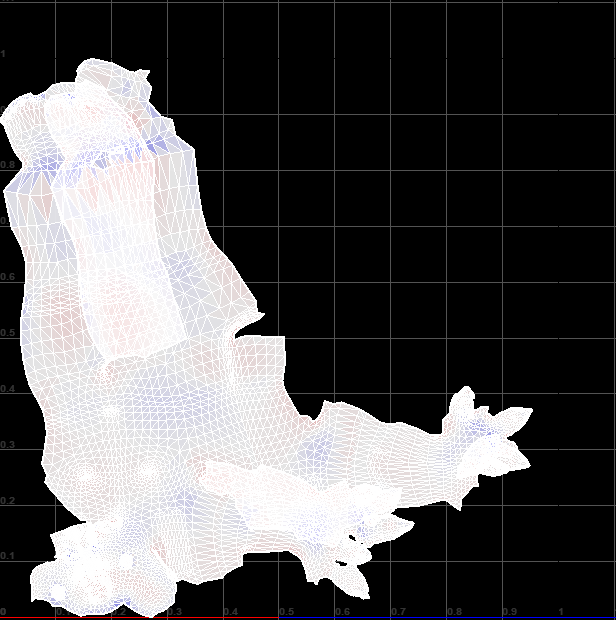}
         \caption{Ken:Optcuts}
         \label{fig:ken_autocuts}
     \end{subfigure}
     \hfill
      \begin{subfigure}[b]{0.15\textwidth}
         \centering
         \includegraphics[width=\textwidth , height=\textwidth]{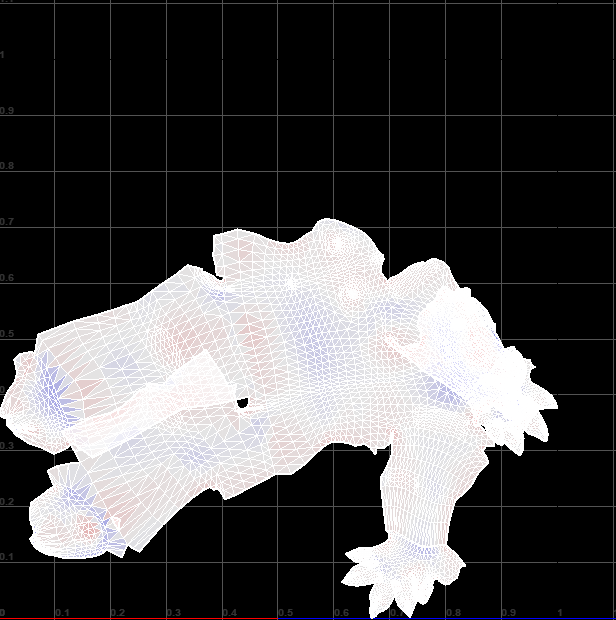}
         \caption{Sibilia:Optcuts}
         \label{fig:sibilia_autocuts}
     \end{subfigure}
     \hfill
      \begin{subfigure}[b]{0.15\textwidth}
         \centering
         \includegraphics[width=\textwidth , height=\textwidth]{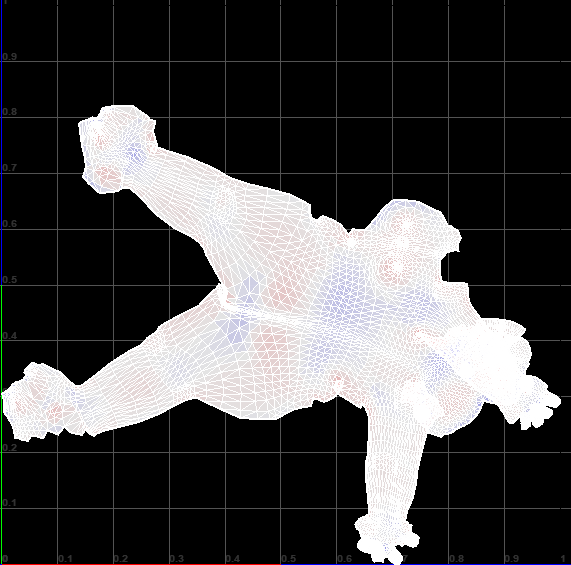}
         \caption{Xena:Optcuts}
         \label{fig:xena_autocuts}
     \end{subfigure}
        \caption{Visualization of the UV maps of the test set of CG10000 using Autocuts (top row) and OptCuts (bottom row) approaches without initialization.}
        \label{fig:autocuts}
\end{figure}

This section provides the UV Mapping visualization (Fig.~\ref{fig:autocuts}) using Autocuts~\cite{poranne2017autocuts} and OptCuts~\cite{li2018optcuts} on our test set objects. As illustrated below, Autocuts and OptCuts only produce a single shell (i.e., without manual assistance provided by a skilled artist) and cannot preserve semantic boundaries.

\section{Autocuts and OptCuts post-processing initialized with GAT output seam labels}
When used during post-processing, Autocuts~\cite{poranne2017autocuts} and OptCuts~\cite{li2018optcuts} can produce UV maps with fewer shells that those derived using our proposed minimum distortion Steiner Tree (DST) algorithm. However, our proposed approach derives results that are closer to the artist-specified ground truth semantics. For example, our approach derives a single shell for the entire upper body while Autocuts and OptCuts break this into multiple shells.

\begin{figure*}[t]
    \centering
    \begin{subfigure}[b]{0.18\textwidth}
        \includegraphics[width=90px, height=90px]{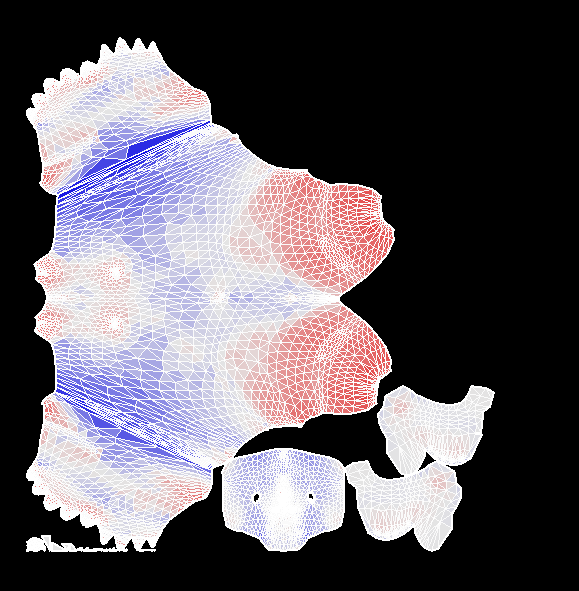}
        \label{fig:gat_ac_initial}
    \end{subfigure}
    \hspace{1px}
    \begin{subfigure}[b]{0.18\textwidth}
        \includegraphics[width=90px, height=90px]{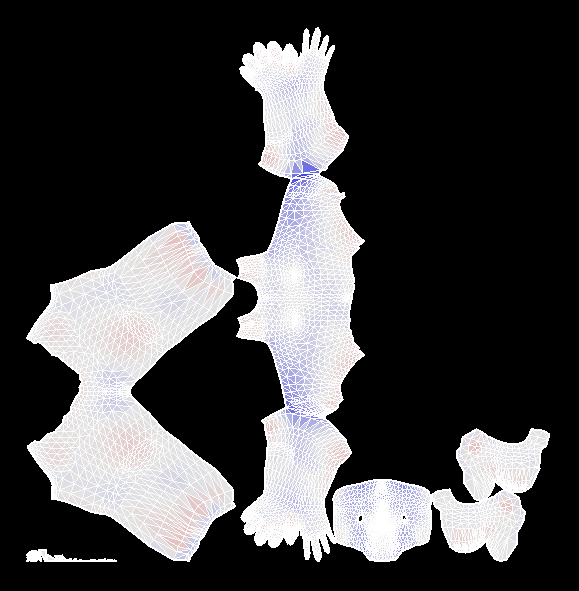}
        \label{fig:gat_ac_25it}
    \end{subfigure}
    \hspace{1px}
    \begin{subfigure}[b]{0.18\textwidth}
        \includegraphics[width=90px, height=90px]{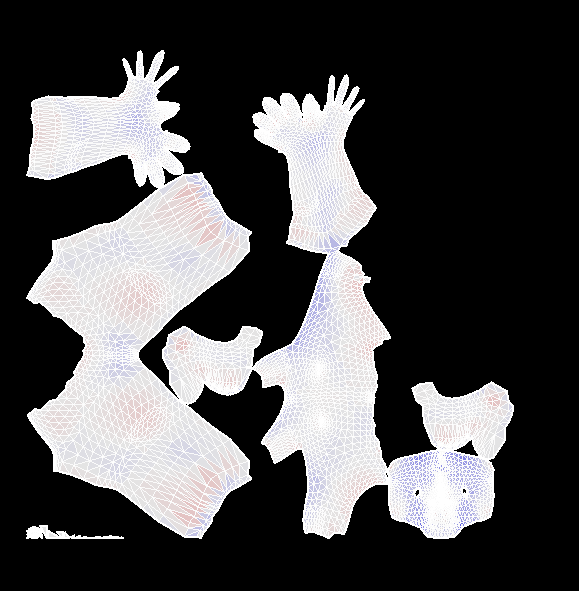}
        \label{fig:gat_ac_50it}
    \end{subfigure}
    \hspace{1px}
    \begin{subfigure}[b]{0.18\textwidth}
        \includegraphics[width=90px, height=90px]{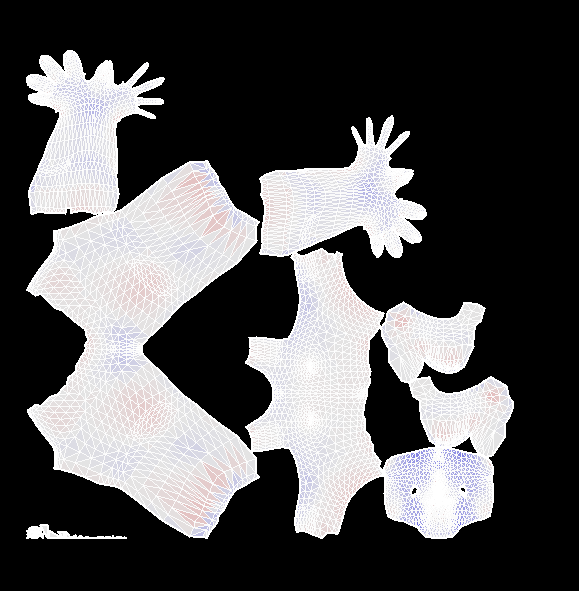}
        \label{fig:gat_ac_100it}
    \end{subfigure}
    \hspace{1px}
    \begin{subfigure}[b]{0.18\textwidth}
        \includegraphics[width=90px, height=90px]{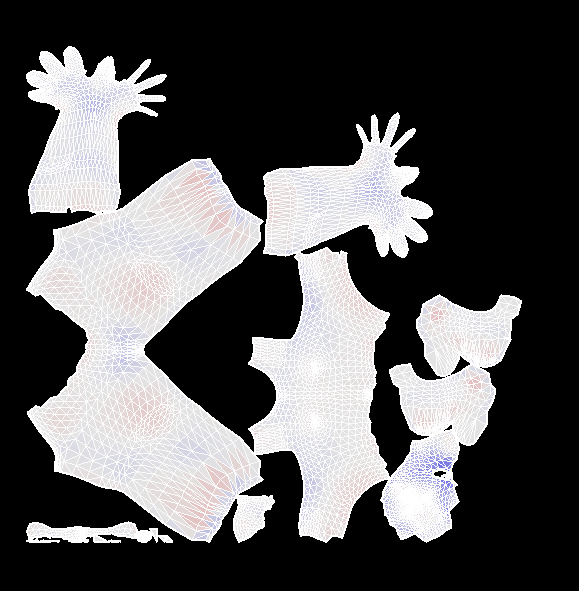}
        \label{fig:gat_ac_200it}
    \end{subfigure}
    \\[-2ex]
    \begin{subfigure}[b]{0.18\textwidth}
        \includegraphics[width=90px, height=90px]{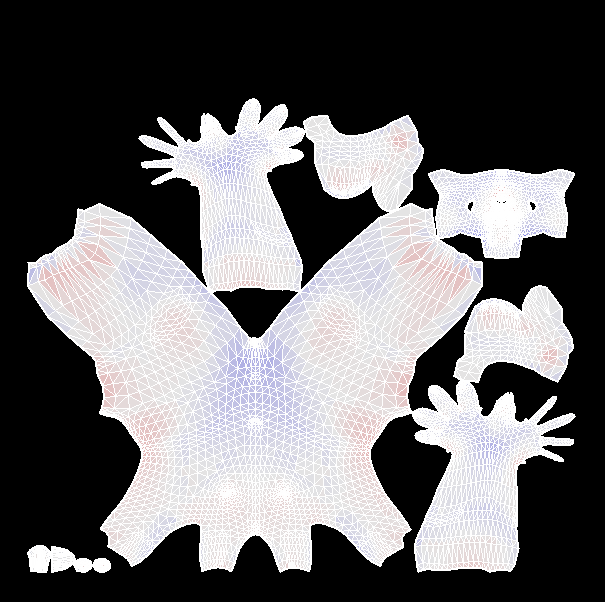}
        \caption{Initial: $\alpha_p=0.9$}
        \label{fig:gat_oc_initial}
    \end{subfigure}
    \hspace{1px}
    \begin{subfigure}[b]{0.18\textwidth}
        \includegraphics[width=90px, height=90px]{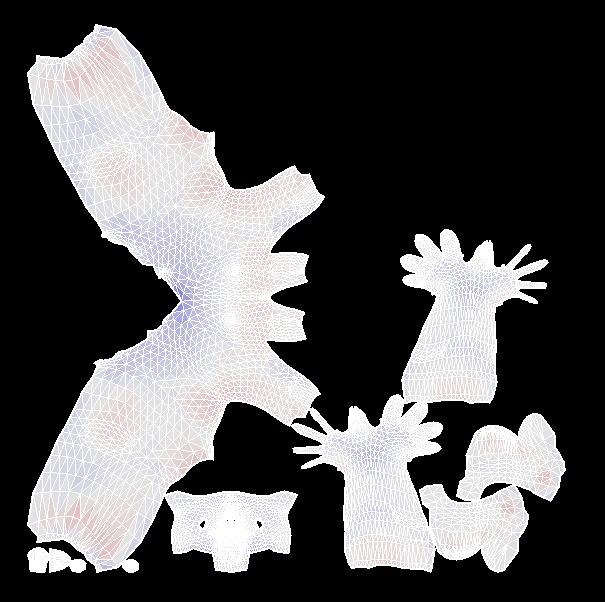}
        \caption{25 iterations}
        \label{fig:gat_oc_25it}
    \end{subfigure}
    \hspace{1px}
    \begin{subfigure}[b]{0.18\textwidth}
        \includegraphics[width=90px, height=90px]{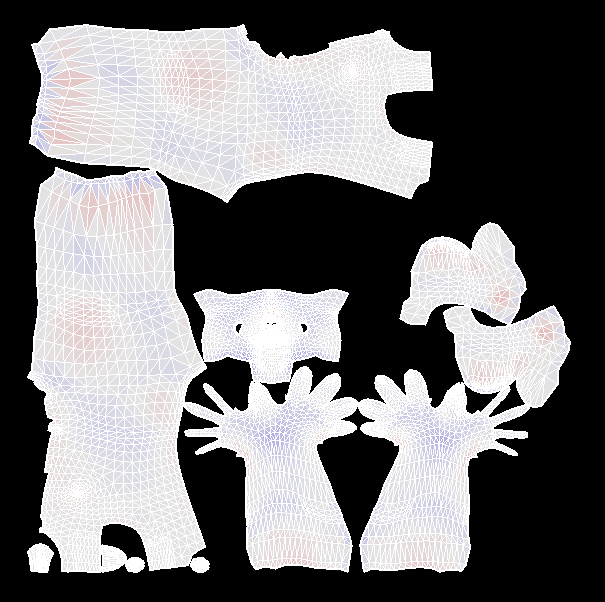}
        \caption{50 iterations}
        \label{fig:gat_oc_50it}
    \end{subfigure}
    \hspace{1px}
    \begin{subfigure}[b]{0.18\textwidth}
        \includegraphics[width=90px, height=90px]{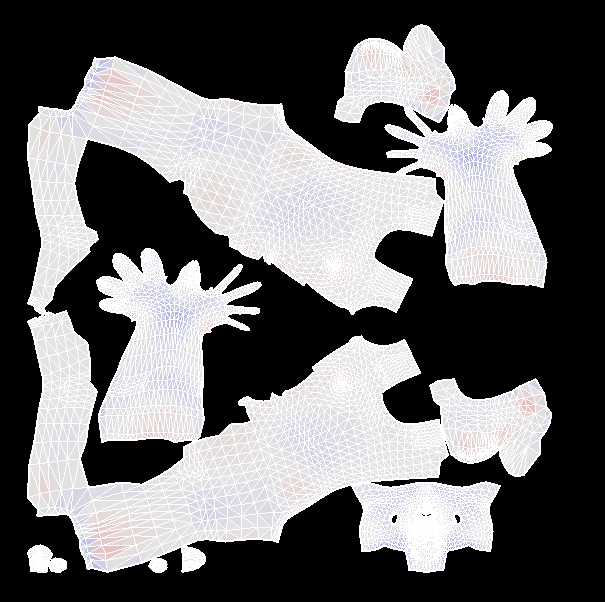}
        \caption{100 iterations}
        \label{fig:gat_oc_100it}
    \end{subfigure}
    \hspace{1px}
    \begin{subfigure}[b]{0.18\textwidth}
        \includegraphics[width=90px, height=90px]{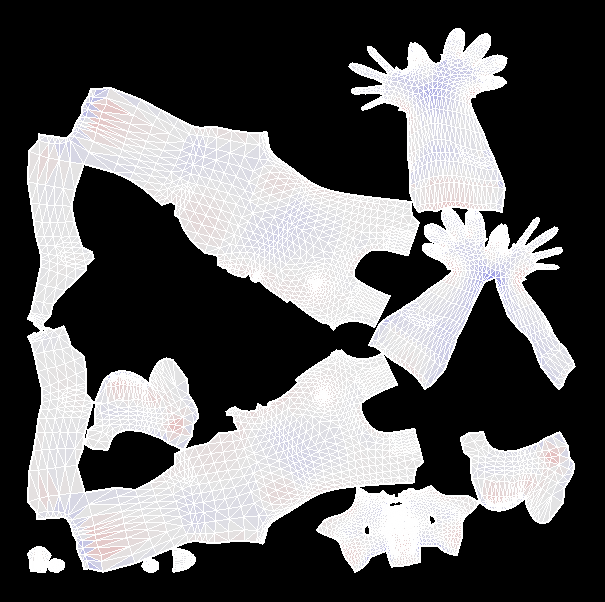}
        \caption{200 iterations}
        \label{fig:gat_oc_200it}
    \end{subfigure}
    \caption{Evolving state of a UV mapping generated using GAT~\cite{velivckovic2017graph} edge probabilities as an initialization step for Autocuts~\cite{poranne2017autocuts} (top row) and OptCuts~\cite{li2018optcuts} (bottom row) approaches on the 3D model Ken.}
    \label{fig:gat_autocuts_optcuts}
\end{figure*}

Figure~\ref{fig:gat_autocuts_optcuts} shows the evolution of the maps produced by post-processing using Autocuts and OptCuts, respectively, when the GAT-based~\cite{velivckovic2017graph} seam detection result is used for initialization.

\section{Visualization of GraphSeam Outputs for the CG10000 test set}\label{CG 10000 visualization}
\begin{figure}[h]
     \centering
     \begin{subfigure}[b]{0.15\textwidth}
         \centering
         \includegraphics[width=\textwidth , height=\textwidth]{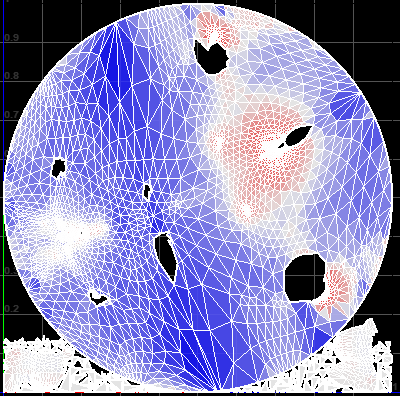}
         \caption{Ken:GCN}
         \label{fig:ken_autocuts}
     \end{subfigure}
     \hfill
     \begin{subfigure}[b]{0.15\textwidth}
         \centering
         \includegraphics[width=\textwidth , height=\textwidth]{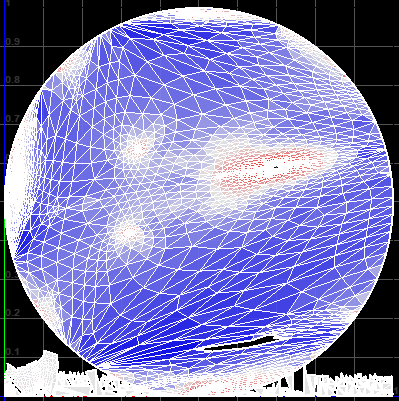}
         \caption{Sibilia:GCN}
         \label{fig:sibilia_autocuts}
     \end{subfigure}
      \hfill
      \begin{subfigure}[b]{0.15\textwidth}
         \centering
         \includegraphics[width=\textwidth , height=\textwidth]{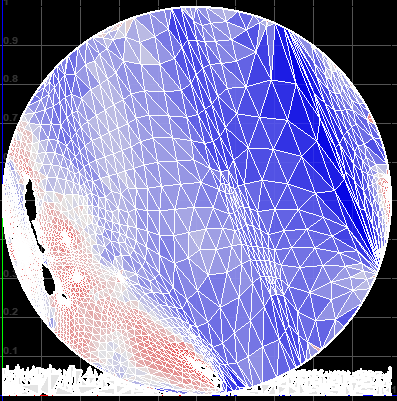}
         \caption{Xena:GCN}
         \label{fig:xena_autocuts}
     \end{subfigure}
     \newline
      \begin{subfigure}[b]{0.15\textwidth}
         \centering
         \includegraphics[width=\textwidth , height=\textwidth]{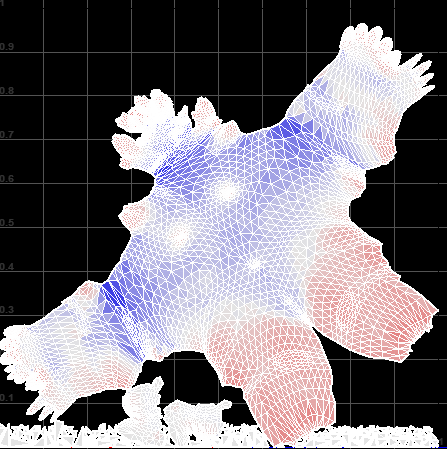}
         \caption{Ken:GCN-DST}
         \label{fig:ken_autocuts}
     \end{subfigure}
     \hfill
      \begin{subfigure}[b]{0.15\textwidth}
         \centering
         \includegraphics[width=\textwidth , height=\textwidth]{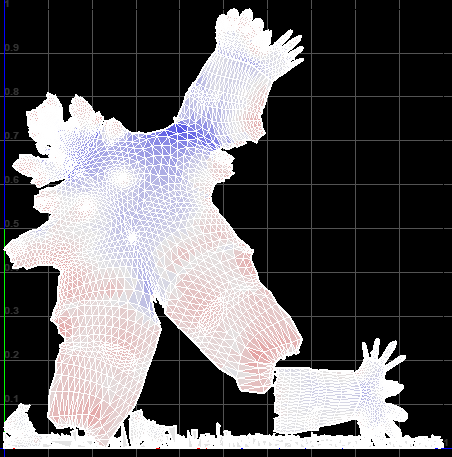}
         \caption{Sibilia:GCN-DST}
         \label{fig:ken_autocuts}
     \end{subfigure}
     \hfill
      \begin{subfigure}[b]{0.15\textwidth}
         \centering
         \includegraphics[width=\textwidth , height=\textwidth]{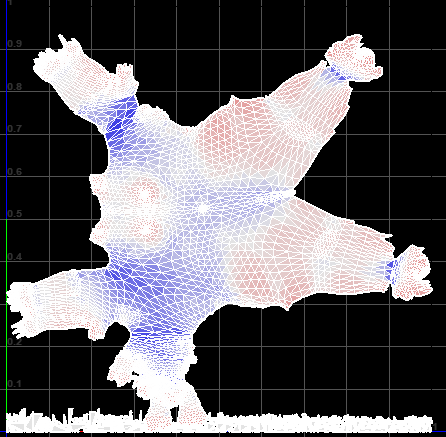}
         \caption{Xena:GCN-DST}
         \label{fig:ken_autocuts}
     \end{subfigure}
     \hfill
        \caption{Visualization of UV maps for GCN~\cite{kipf2016semi} (top row) and GCN-DST (bottom row) on CG10000 test objects.}
        \label{fig:gcn_steiner}
\end{figure}

\begin{figure}[h]
     \centering
          \begin{subfigure}[b]{0.15\textwidth}
         \centering
         \includegraphics[width=\textwidth , height=\textwidth]{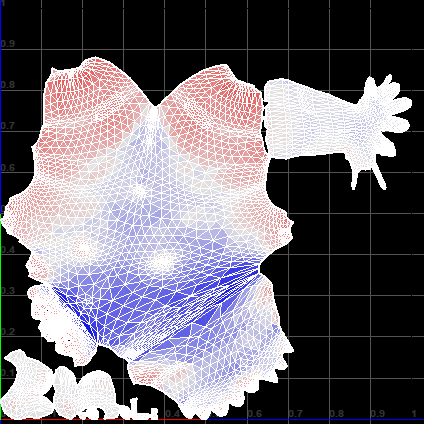}
         \caption{Ken:{\tiny GCN-SK}}
         \label{fig:ken_autocuts}
     \end{subfigure}
     \hfill
          \begin{subfigure}[b]{0.15\textwidth}
         \centering
         \includegraphics[width=\textwidth , height=\textwidth]{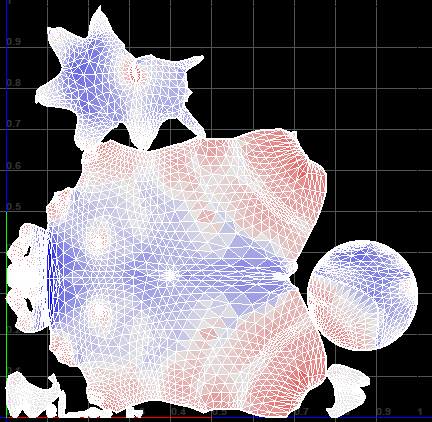}
         \caption{Sibilia:{\tiny GCN-SK}}
         \label{fig:ken_autocuts}
     \end{subfigure}
      \hfill
          \begin{subfigure}[b]{0.15\textwidth}
         \centering
         \includegraphics[width=\textwidth , height=\textwidth]{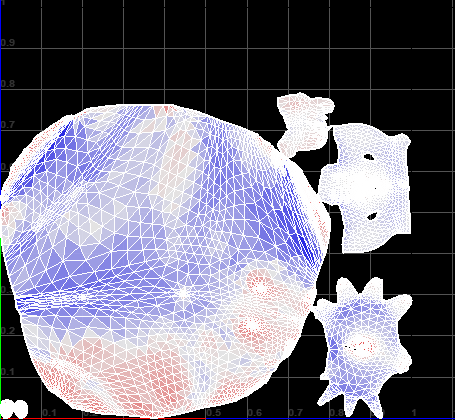}
         \caption{ Xena:{\tiny GCN-SK}}
         \label{fig:ken_autocuts}
     \end{subfigure}
      \newline
      \begin{subfigure}[b]{0.15\textwidth}
         \centering
         \includegraphics[width=\textwidth , height=\textwidth]{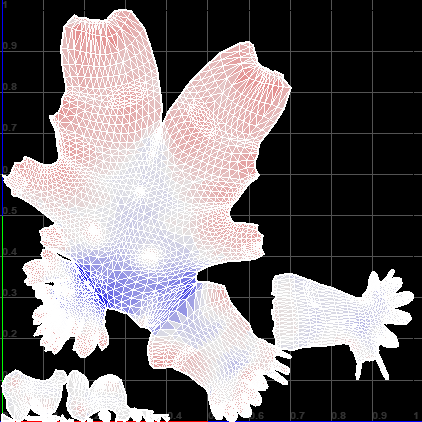}
         \caption{Ken:{\tiny GCN-SK-DST}}
         \label{fig:ken_autocuts}
     \end{subfigure}
     \hfill
     \begin{subfigure}[b]{0.15\textwidth}
         \centering
         \includegraphics[width=\textwidth , height=\textwidth]{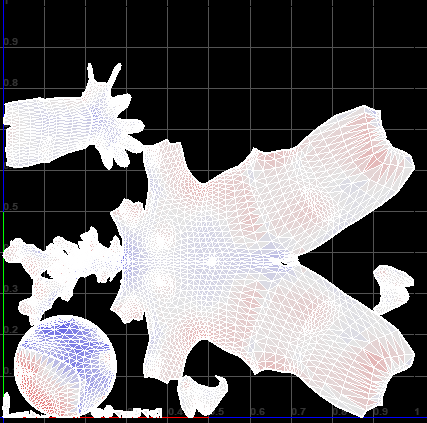}
         \caption{{\tiny Sibilia: GCN-SK-DST}}
         \label{fig:ken_autocuts}
     \end{subfigure}
    \hfill
          \begin{subfigure}[b]{0.15\textwidth}
         \centering
         \includegraphics[width=\textwidth , height=\textwidth]{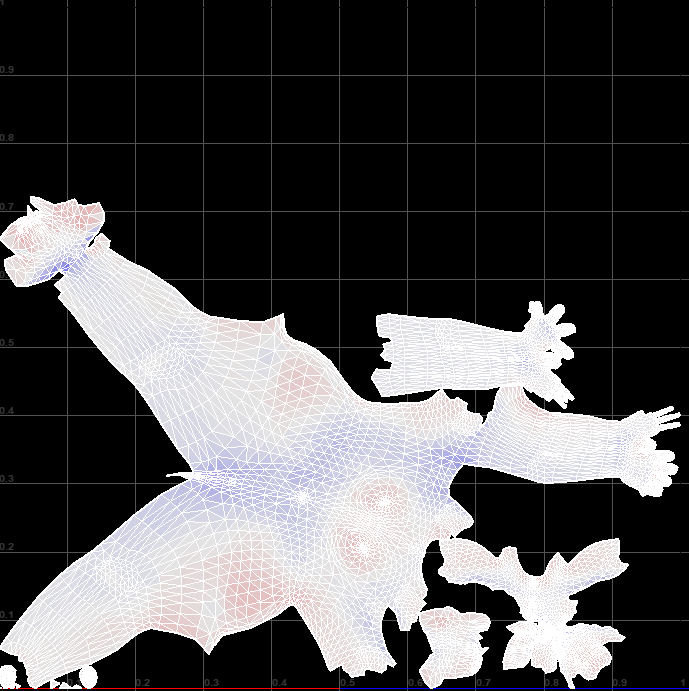}
         \caption{Xena:{\tiny GCN-SK-DST}}
         \label{fig:ken_autocuts}
     \end{subfigure}
    \hfill
    \caption{Visualization of UV maps for GCN-SK (top row) and GCN-SK-DST (bottom row) on CG10000 test objects.}
        \label{fig:gcn_skele_steiner}
\end{figure}

\begin{figure}[h]
     \centering
     \begin{subfigure}[b]{0.15\textwidth}
         \centering
         \includegraphics[width=\textwidth , height=\textwidth]{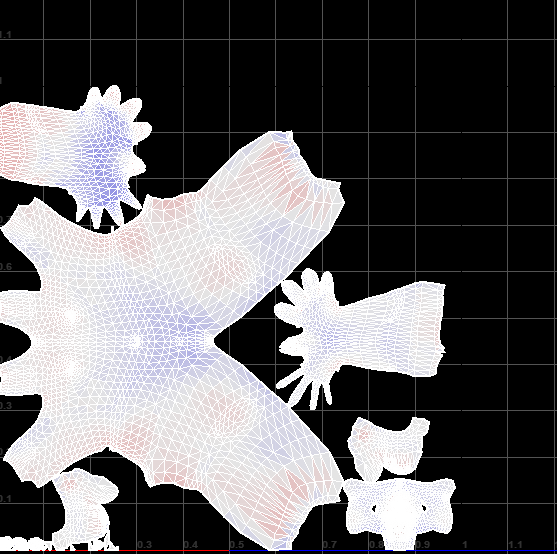}
         \caption{Ken:GAT}
         \label{fig:ken_autocuts}
     \end{subfigure}
     \hfill
     \begin{subfigure}[b]{0.15\textwidth}
         \centering
         \includegraphics[width=\textwidth , height=\textwidth]{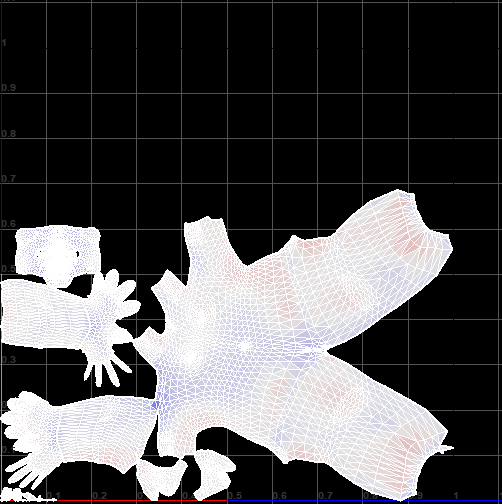}
         \caption{Sibilia:GAT}
         \label{fig:sibilia_autocuts}
     \end{subfigure}
     \hfill
     \begin{subfigure}[b]{0.15\textwidth}
         \centering
         \includegraphics[width=\textwidth , height=\textwidth]{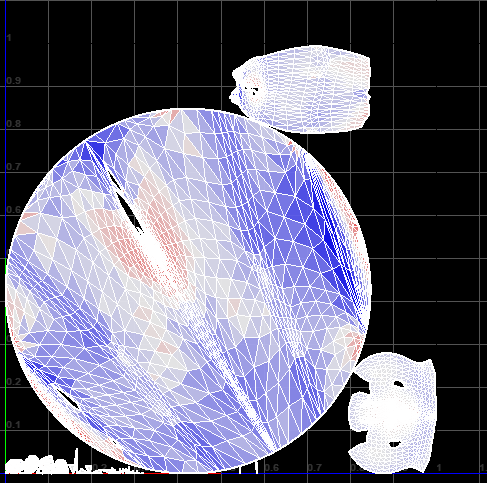}
         \caption{Xena:GAT}
         \label{fig:xena_autocuts}
     \end{subfigure}
      \newline
      \begin{subfigure}[b]{0.15\textwidth}
         \centering
         \includegraphics[width=\textwidth , height=\textwidth]{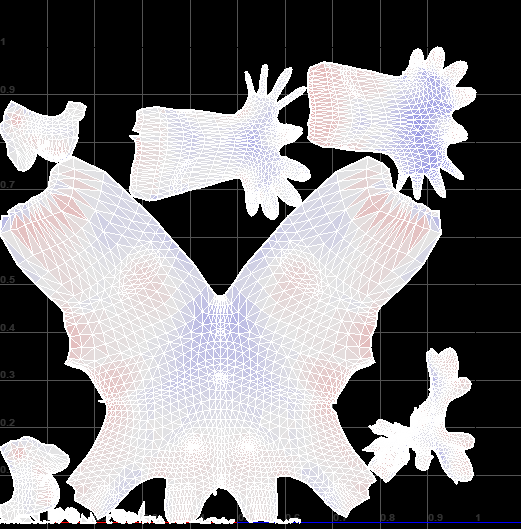}
         \caption{Ken:GAT-DST}
         \label{fig:ken_autocuts}
     \end{subfigure}
      \hfill
      \begin{subfigure}[b]{0.15\textwidth}
         \centering
         \includegraphics[width=\textwidth , height=\textwidth]{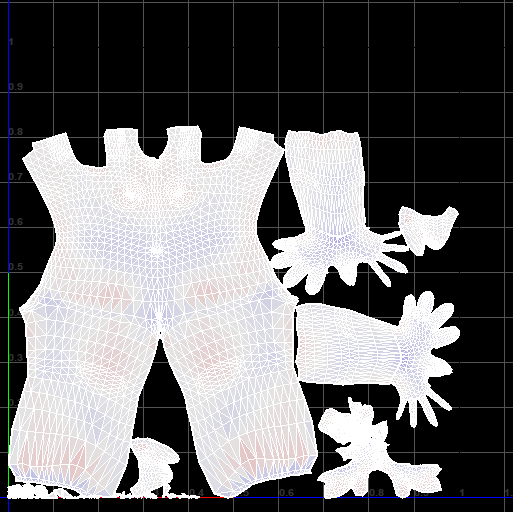}
         \caption{Sibilia:{\tiny GAT-DST}}
         \label{fig:ken_autocuts}
     \end{subfigure}
     \hfill
          \begin{subfigure}[b]{0.15\textwidth}
         \centering
         \includegraphics[width=\textwidth , height=\textwidth]{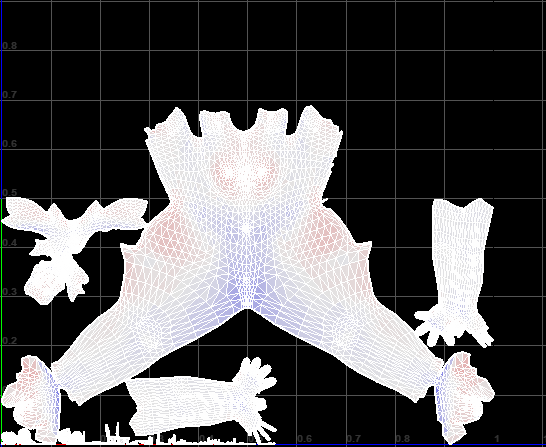}
         \caption{Xena:GAT-DST}
         \label{fig:ken_autocuts}
     \end{subfigure}
     \hfill
        \caption{Visualization of UV maps for GAT\cite{velivckovic2017graph} (top row) and GAT-DST (bottom row) on CG10000 test objects.}
        \label{fig:gat}
\end{figure}

\begin{figure}[h]
     \centering
     \begin{subfigure}[b]{0.15\textwidth}
         \centering
         \includegraphics[width=\textwidth , height=\textwidth]{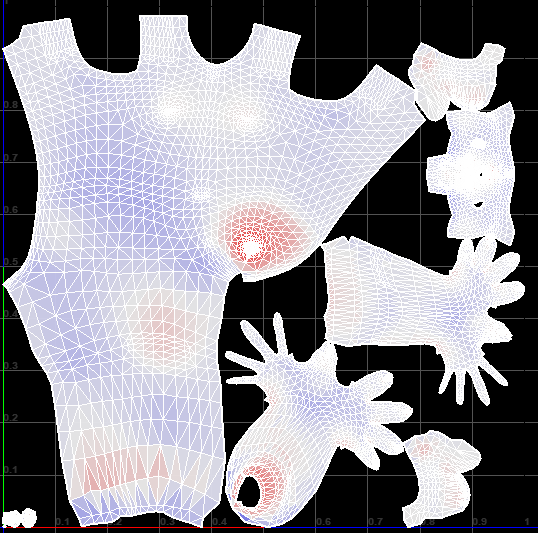}
         \caption{Ken:GS}
         \label{fig:ken_autocuts}
     \end{subfigure}
     \hfill
     \begin{subfigure}[b]{0.15\textwidth}
         \centering
         \includegraphics[width=\textwidth , height=\textwidth]{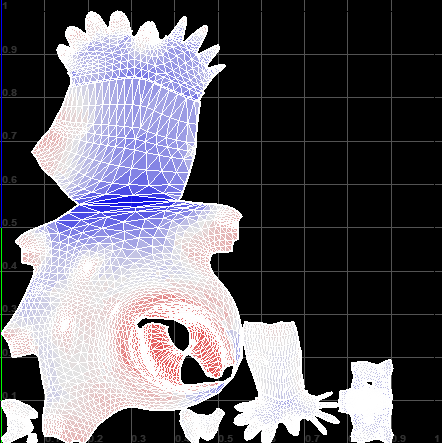}
         \caption{Sibilia:GS}
         \label{fig:sibilia_autocuts}
     \end{subfigure}
     \hfill
     \begin{subfigure}[b]{0.15\textwidth}
         \centering
         \includegraphics[width=\textwidth , height=\textwidth]{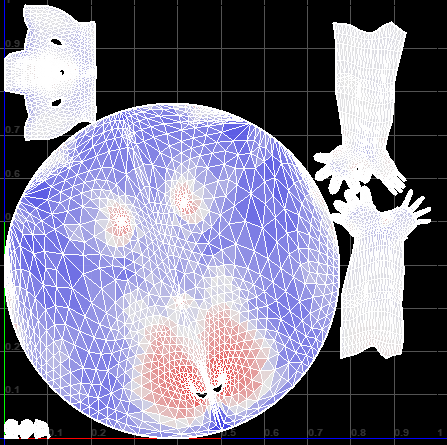}
         \caption{Xena:GS}
         \label{fig:xena_autocuts}
     \end{subfigure}
      \hfill
      \begin{subfigure}[b]{0.15\textwidth}
         \centering
         \includegraphics[width=\textwidth , height=\textwidth]{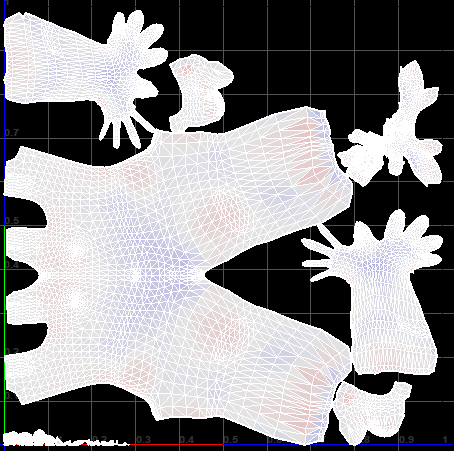}
         \caption{Ken:GS-DST}
         \label{fig:ken_autocuts}
     \end{subfigure}
      \hfill
      \begin{subfigure}[b]{0.15\textwidth}
         \centering
         \includegraphics[width=\textwidth , height=\textwidth]{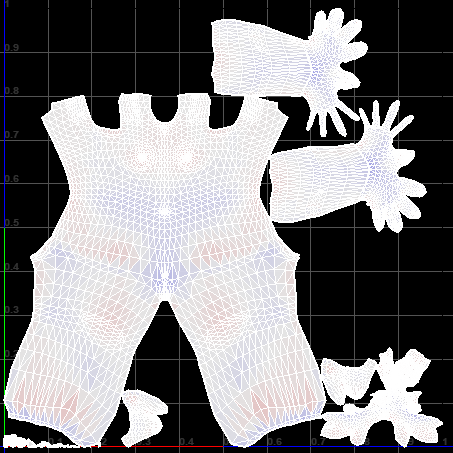}
         \caption{Sibilia:GS-DST}
         \label{fig:ken_autocuts}
     \end{subfigure}
     \hfill
          \begin{subfigure}[b]{0.15\textwidth}
         \centering
         \includegraphics[width=\textwidth , height=\textwidth]{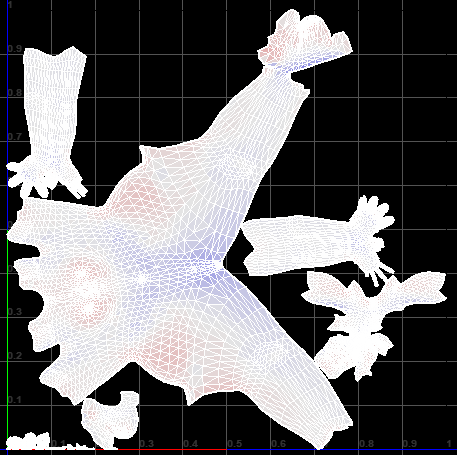}
         \caption{Xena:GS-DST}
         \label{fig:ken_autocuts}
     \end{subfigure}
        \caption{Visualization of UV maps for GraphSAGE\cite{hamilton2017inductive} and GraphSAGE-DST on CG10000 test objects.}
        \label{fig:graphsage}
\end{figure}

\begin{figure}[h]
     \centering
     \begin{subfigure}[b]{0.15\textwidth}
         \centering
         \includegraphics[width=\textwidth , height=\textwidth]{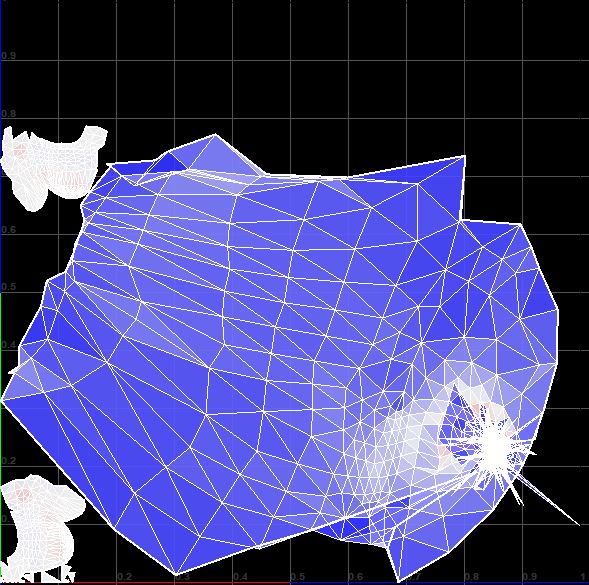}
         \caption{Ken:GIN}
         \label{fig:ken_gin}
     \end{subfigure}
     \hfill
     \begin{subfigure}[b]{0.15\textwidth}
         \centering
         \includegraphics[width=\textwidth , height=\textwidth]{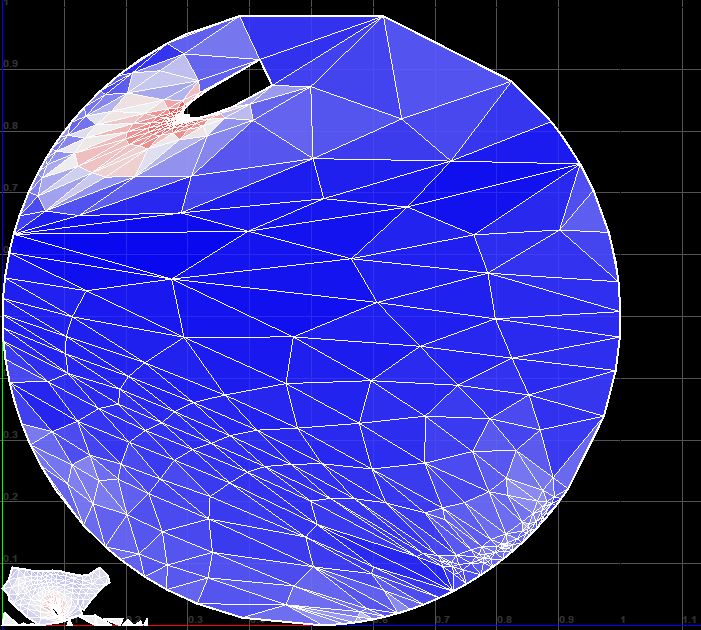}
         \caption{Sibilia:GIN}
         \label{fig:sibilia_gin}
     \end{subfigure}
     \hfill
     \begin{subfigure}[b]{0.15\textwidth}
         \centering
         \includegraphics[width=\textwidth , height=\textwidth]{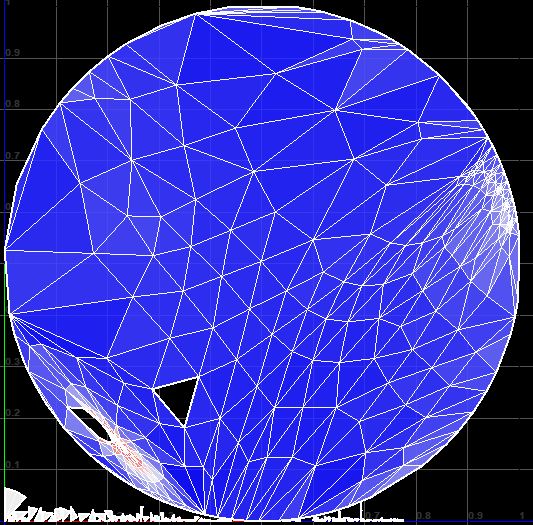}
         \caption{Xena:GIN}
         \label{fig:xena_gin}
     \end{subfigure}
      \newline
      \begin{subfigure}[b]{0.15\textwidth}
         \centering
         \includegraphics[width=\textwidth , height=\textwidth]{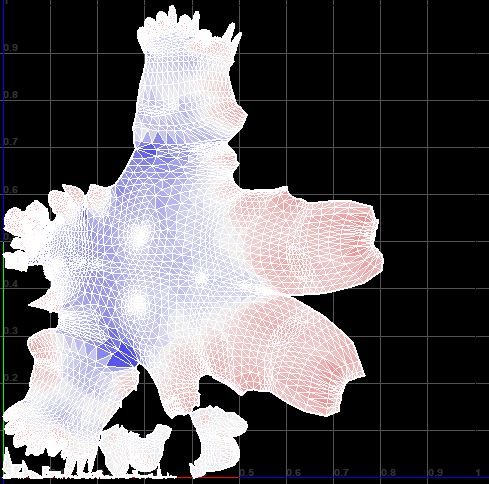}
         \caption{Ken:GIN-DST}
         \label{fig:ken_autocuts}
     \end{subfigure}
      \hfill
     \begin{subfigure}[b]{0.15\textwidth}
         \centering
         \includegraphics[width=\textwidth , height=\textwidth]{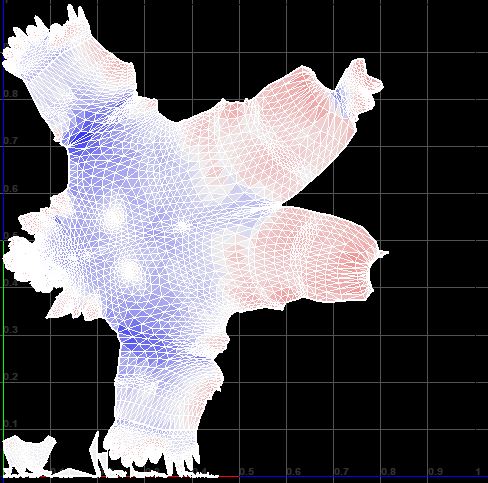}
         \caption{{\tiny Sibilia:GIN-DST}}
         \label{fig:ken_autocuts}
     \end{subfigure}
     \hfill
     \begin{subfigure}[b]{0.15\textwidth}
         \centering
         \includegraphics[width=\textwidth , height=\textwidth]{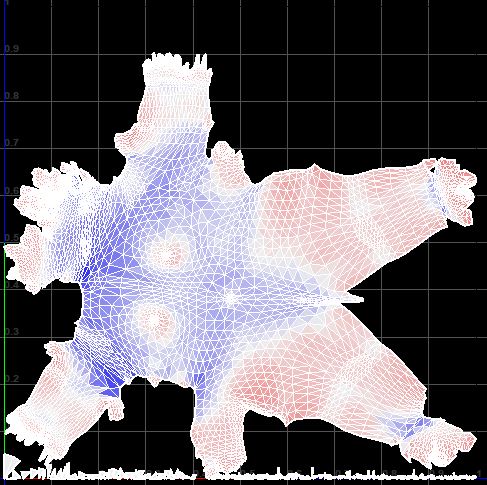}
         \caption{Xena:GIN-DST}
         \label{fig:ken_autocuts}
     \end{subfigure}
        \caption{Visualization of UV maps for GIN\cite{xu2018powerful} and GIN-DST on CG10000 test objects.}
        \label{fig:gin}
\end{figure}

In this section, we provide the results for all the objects in the test set for our proposed method, GraphSeam, using GNNs including GCN, GAT, GraphSAGE and GIN for the seam detection block. The visualizations are derived after applying of our suggested post-processing approach, Distortion Steiner Tree (DST). Figures~\ref{fig:gcn_steiner} to \ref{fig:gin} display visualizations of the derived UV maps.

Based on the results, it is clear that the attention mechanism in GAT and the more powerful aggregation method in GraphSAGE (we use the LSTM aggregator) play a critical role in producing good results for seam classification. GCN, which uses only a simple averaging aggregator, produces much poorer initial UV maps. GIN is known to perform excellently for graph classification and captures the graph structure as well as most other GNNs. Its performance for node or edge classification tasks can be poorer, because it can be important to place more emphasis on local features. This probably explains the considerably weaker performance for the seam detection task, which we formulate as edge classification. 

All of the results clearly show how applying DST to the outputs of the seam detection block dramatically improves the final UV maps and leads to better preservation of semantic boundaries.

Figure~\ref{fig:gcn_skele_steiner} provides UV map visualizations that show the result of applying the proposed skeletonization(SK) procedure on the GCN output. Comparing to Fig.~\ref{fig:gcn_steiner}, we see that by thinning the edges there is a dramatic reduction in the number of small shells. The single large shell produced by GCN is decomposed into more meaningful shells, providing a better initialization for the DST algorithm.

\newpage
\section{Robustness of our method}
\subsection{Using random splits}
In the main paper we provide our experimental results for a single validation/test split. To explore robustness to different data splits, we conducted experiments using five (5) random validation and test splits. The results are reported in Table~\ref{table:CG_10000_pef_metrics_random_splits}. We observe that although there is some variability in performance, our proposed seam detection method achieves high accuracy and low FPR for all splits when using GAT and GraphSAGE.

\subsection{Using augmentation on datasets}
We have developed two tools for artists to create datasets that can be used to train the GraphSeam method. We employed the decimation tool to produce meshes with 10000 faces resolution for the CG10000 dataset.

In this section, we explore the use of the augmentation tool to produce a dataset that is twice the size of the original CG dataset. The augmentation tool is based on adding random Gaussian noise to the original vertex positions. For this set of experiments, the amount of noise varies vertex positions up to 20\%, leading to a much more varied dataset and making the seam detection and mapping tasks harder.

\begin{table}[h]
\centering
\caption{Performance of seam detection using false positive rate (FPR), true positive rate (TPR) and accuracy (Acc.) on CG10000 with augmentation.}
\vspace{0.1cm}
\label{table:CG_10000_pef_metrics_augment}
\begin{tabular}{|p{2.3cm}|p{0.5cm}|p{0.5cm}|p{0.5cm}|}
 \hline
  \multirow{2}{*}{\textbf{Method}} & \multicolumn{3}{c|}{\textbf{Perf. Metrics (\%)}} \\ \cline{2-4}
  & \textbf{FPR} & \textbf{TPR} & \textbf{Acc.}\\
 \hline
 \hline
    Prop-GCN &3.30& 94.83& 96.62 \\
 \hline    
  Prop-GAT &0.17& 99.08& 99.80 \\
  \hline
  Prop-GS (pool) &0.12& 97.01& 99.76\\
  \hline
    Prop-GS (mean)&1.13&97.57&98.81\\
  \hline
    Prop-GS (GCN) &4.64&95.80&95.37\\
  \hline
    Prop-GS (LSTM) &0.03&98.20&99.89\\
  \hline
  Prop-GIN &1.00& 82.06& 98.36 \\
 \hline
\end{tabular}
\end{table}

To provide a fair comparison, we are using the same split for training, validation and test set as we reported in the paper and we use the augmentation method on each set separately to ensure there is no overlap between splits.

Table~\ref{table:CG_10000_pef_metrics_augment} reports the results of the GraphSeam seam detection algorithm for the augmented dataset. Despite the increased variability in the dataset, the results are very similar to those achieved for the original CG10000 dataset. This carries through to the derived UV maps after post-processing, so we do not show the results here.

\subsection{Creating stylized datasets}
To show the power of the augmentation tool, we build our new dataset using 19 random original 3D models from the initial training set of CG10000 and via augmentation we increase the new training set to 93 (the same size as the original CG10000 training set). To provide a fair comparison we are using the same validation and test set as CG10000. It is required to mention that we did not make use of the augmentation tool in the main paper, because professional artists had already provided a sufficient number of labelled 3D models.

Our goal in providing this set of results is to show that our proposed augmentation method enables artists to create their dataset based on a few original models containing manual seams using the augmentation tool. As shown in Table~\ref{table:CG_10000_pef_metrics_fewer_models}, seam detection results illustrate similar performance to that obtained with the original CG10000 dataset.

\begin{table}[h]
\centering
\begin{tabular}{|p{2.3cm}|p{0.5cm}|p{0.6cm}|p{0.5cm}|}
 \hline
  \multirow{2}{*}{\textbf{Method}} & \multicolumn{3}{c|}{\textbf{Perf. Metrics (\%)}} \\ \cline{2-4}
  & \textbf{FPR} & \textbf{TPR} & \textbf{Acc.}\\
 \hline
 \hline
    Prop-GCN &7.89& 92.54& 92.13\\
 \hline    
  Prop-GAT &0.57&97.57& 99.35 \\
  \hline
  Prop-GS (pool) &0.75&92.43& 98.97\\
  \hline
    Prop-GS (mean)&1.21& 92.88&98.55\\
  \hline
    Prop-GS (GCN) &10.25& 92.99&89.88\\
  \hline
    Prop-GS (LSTM) &0.10& 93.76& 99.64 \\
  \hline
  Prop-GIN &2.41& 82.33& 97.51 \\
 \hline
\end{tabular}
\caption{Performance of seam detection using false positive rate (FPR), true positive rate (TPR) and accuracy (Acc.) on CG10000 containing fewer train models.}
\vspace{0.1cm}
\label{table:CG_10000_pef_metrics_fewer_models}
\end{table}

\begin{table*}[t]
\centering
\caption{Performance of seam detection is evaluated using false positive rate (FPR), true positive rate (TPR) and accuracy (Acc.) on different random validation and test splits.}
\vspace{0.1cm}
\label{table:CG_10000_pef_metrics_random_splits}
\footnotesize
\begin{tabular}{|p{2cm}|p{0.5cm}|p{0.5cm}|p{0.5cm}|p{0.5cm}|p{0.5cm}|p{0.5cm}|p{0.7cm}|p{0.7cm}|p{0.7cm}|p{0.5cm}|p{0.5cm}|p{0.5cm}|}
 \hline
  \multirow{2}{*}{\textbf{Split Method}} & \multicolumn{3}{c|}{\textbf{GCN}} & \multicolumn{3}{c|}{\textbf{GAT}} & \multicolumn{3}{c|}{\textbf{GraphSAGE (LSTM)}} &
  \multicolumn{3}{c|}{\textbf{GIN}} \\ \cline{2-13}
  & \textbf{FPR} & \textbf{TPR} & \textbf{Acc.} & \textbf{FPR} & \textbf{TPR} & \textbf{Acc.} & \textbf{FPR} & \textbf{TPR} & \textbf{Acc.} & \textbf{FPR} & \textbf{TPR} & \textbf{Acc.}\\
 \hline
 \hline
 Random 1 &3.62& 87.36& 96.01& 0.34& 99& 99.63& 0.04& 98.56& 99.9& 0.85& 71.74& 98.05\\
 \hline
 Random 2& 5.11& 90.22& 94.70& 0.39 & 99.74& 99.52& 0.56& 98.22& 99.39& 5.77& 92	&95.94\\
 \hline
  Random 3 &4.07& 90.69& 95.72& 0.62& 98.78& 99.35& 0.083& 98& 99.84& 1.89& 86.93&97.95\\
 \hline
  Random 4 &3.83& 96.72& 96.18& 0.37& 98.94& 99.6& 0.04& 99.05& 99.92& 1.2& 91.45& 98.68\\
 \hline
  Random 5 &8.83& 95.72& 91.35& 0.67& 97.5& 99.26& 0.08& 95.67& 99.75& 3.47& 90.12& 97.21\\
 \hline
  Average &5.09& 92.14& 94.79& 0.47& 98.79& 99.47& 0.16& 97.90 &99.76& 2.63& 86.44 & 97.56\\
 \hline
\end{tabular}
\end{table*}

\end{document}